\documentclass[9pt]{article}
\usepackage[super,sort&compress,comma]{natbib} 
\usepackage{amsmath}
\usepackage{graphicx} 
\usepackage{epstopdf}

\begin{document}

\title{{\textbf{Photonics and Spectroscopy in Nanojunctions: A Theoretical Insight}}}

\author{\Large{Michael Galperin}}

\date{January 27, 2017}
\maketitle

\begin{abstract}
Progress of experimental techniques at nanoscale in the last decade  made optical 
measurements in current-carrying nanojunctions a reality thus indicating emergence of a new
field of research coined as optoelectronics.
Optical spectroscopy of open nonequilibrium systems is a natural meeting point for
(at least) two research areas: nonlinear optical spectroscopy and quantum transport,
each with its own theoretical toolbox. We review recent progress in
the field comparing theoretical treatments of optical response in nanojunctions
as is accepted in nonlinear spectroscopy and quantum transport communities.
A unified theoretical description of spectroscopy in nanojunctions is presented.
We argue that theoretical approaches of the quantum transport community 
(and in particular, the Green function based considerations) yield a convenient
tool for optoelectronics when radiation field is treated classically, 
and that differences between the toolboxes may become critical 
when studying quantum radiation field in junctions.
\end{abstract}


\footnotetext{\textit{Department of Chemistry \& Biochemistry, University of California San Diego, 9500 Gilman Dr., La Jolla, CA 92093, USA. Tel: +1 858 246 0511; E-mail: migalperin@ucsd.edu}}

\section{Introduction}\label{intro}
Optical spectroscopy is an important diagnostic tool routinely applied to study molecules 
(either in gas, or condensed phases, or adsorbed on surfaces).
In nanojunctions spectroscopic applications range from characterization of molecular structures and
junction interfaces, to introducing nanoscale thermometry, to inducing and  controlling 
molecular dynamics and chemistry. 
Wide range of spectroscopic techniques is utilized in the studies including (to name a few) infrared,\cite{ZhuJACS04,VardenySolStComm05,SchonenbergerJPCC07,ScottJPCC08} 
X-ray,\cite{OckoPNAS06,ConociAdvFuncMat06} 
sum frequency generation (SFG),\cite{DlottScience07,DlottAccChemRes09,UosakiYuJPCC09,DaviesJPCL12} 
as well as surface- (SERS)\cite{NitzanScience07,YuRenJPCC08,MayorWandlowskiACSNano11,VanDuynePCCP13,VanDuyneJPCC15}
and tip-enhances (TERS)\cite{VanDuyneNL12,VanDuyneJPCC12,VanDuyneJPCL14,VanDuyneJACS14,SeidemanVanDuyneNL15,RenAnalytChem16,ToccfondiJPCC16}
Raman spectroscopies.

In recent years optical experiments in current-carrying single-molecule junctions became
a reality.\cite{TianJACS06,NatelsonJPCM08,SchneiderSurfSciRep10,SelzerChemSocRev11,NatelsonPCCP13,VenkataramanNatNano13,KimFrontPhys14,KiguchiPCCP15} 
Combination of the fields of optical spectroscopy and molecular electronics indicates emergence of 
a new field of research, coined molecular optoelectronics.\cite{MGANPCCP12}
In particular, mutliple experiments reported detection of current-induced photon emission 
(electroluminiscence).\cite{CallantApplPhysLett,BerndtPRB02,BerndtPRL03,SchneiderPRL05,YanagiSolidFilms06,HsuJApplPhys07,BrodardCHIMIA07,HoPRB08,BerndtPRL09,BerndtPRL10,HoPRL10,DongNatPhoton10,BerndtPRB11,BerndtPRB12,BerndtPRL12,LothAPL13,ApkarianACSNano14,BerndtSurfSci16,HoJPCC16,MiwaNature16} 
Among them vibrationally\cite{HoPRB08,HoJPCC16} and spatially\cite{HoPRL10}
resolved photo-emission, electroluminiscence as a measure of multi-electron 
processes\cite{BerndtPRL10,DongNatPhoton10,BerndtPRL12,BerndtSurfSci16}
and noise\cite{BerndtPRB11,BerndtPRL12,BerndtPRB12} in junctions, 
as well as indicator of vibronic motion\cite{ApkarianACSNano14} 
and  real space energy transfer\cite{MiwaNature16} were reported.
Alternatively, external illumination was utilized as means to control electron transfer
and transport.\cite{VanDerMolen09,SelzerAngChem10,RalphACSNano11,LindsayNL11,CaudaRSCAdv12,SelzerJPCL13,VanDuyneJPCL15,ErbeAdvSci15,VenkataramanPCCP16,FrisbieJPCC16,ApkarianWuJPCC16} 
Measurements of light induced magnetization in chiral molecules were also 
reported in the literature.\cite{NaamanPRL10,NaamanScience11,NaamanWaldeckJPCL12,PaltielNL14,NaamanNL15,NaamanNL16}

For molecules chemisorbed on metallic surfaces or encapsulated in nanocavities
molecular excitations are coupled with plasmons. This interaction leads to enhancement of molecular
signal,\cite{VanDuyne1977,NieEmoryScience97,FeldPRL97,TianJACS11} which yields possibility to measure optical response of single molecule.\cite{GerstenBirkeLombardiPRL79,GerstenNitzanJCP80,PerssonCPL81,VanDuyne2007}
Thus engineering effective plasmonic structures is crucially important,
and nanoplamonics becomes an inherent part in construction and operation
of any molecular optoelectronic device.\cite{NatelsonNL07,SchonenbergerJPCC07,NordlanderHalasNL08,BerndtPRL09,NordlanderAizpuruaNL10,DongNatPhoton10,SelzerNL11,BerndtPRB12,SelzerJPCL13,NatelsonNL13,VanDuyneJPCL13,NordlanderHalasNatCommun14,BaumbergNL15,VanDuyneJPCL15,NatelsonJPCC16,PotmaApkarianJPCC16}
Note that while usually construction of nanometer scale gaps is the way to form areas 
of high electromagnetic field (hot spots),\cite{NatelsonNL07} possibility of surface enhanced 
spectroscopy without nanogaps was also reported.\cite{VanDuyneJPCL13}

As already mentioned surface enhanced Raman (SERS)\cite{MoskovitsCPL08,TianPCCP10,TianNature10,MoskovitsPCCP13} 
and tip-enhanced Raman (TERS)\cite{VanDuyneJPCL14review,TianNatRevMat16} spectroscopies 
are utilized as indicators of structural changes and dynamics in junctions.
Simultaneous measurements of SERS and conductivity\cite{ZouAnalytChem06,CheshnovskySelzerNatNano08,NatelsonNL08,NatelsonNatNano11,StoddartMirkinSmall13,KiguchiJPCC13,KronikNeatonNatelsonPNAS14,KiguchiIntJModPhysB16,KronikNeatonNatelsonNL16}
provide information on dynamical correlations between the two signals\cite{NatelsonNL08,BanikApkarianParkMGJPCL13}
and current-induced heating,\cite{CheshnovskySelzerNatNano08,NatelsonNatNano11} 
and characterize charging states of molecules in junctions.\cite{WhiteTretiakNL14,KronikNeatonNatelsonPNAS14,KronikNeatonNatelsonNL16}

While initially most of optical experiments in junctions were focused on steady-state response, 
lately time-dependent and transient characteristics started to attract attention.
Optical pump-probe type measurements in junctions were realized in the form of 
time-dependent voltage induced plasmonic luminiscence.\cite{LothAPL13}
Laser pulse induced transport measurements as a tool to asses the intramolecular
dynamics on a sub-picosecond time scale was also suggested.\cite{SelzerPeskinJPCC13,OchoaSelzerPeskinMGJPCL15}
Recently multidimensional spectroscopy measurements in presence of current (although not yet in junctions) 
were reported in the literature.\cite{CundiffOptExpres13,MarcusNatComm14,BittnerSciRep16}

Quantum effects in light-matter interactions is another recent development. 
For example, quantum effects in nanoplasmonics (such as transition of entanglement
between photons and plasmons) recently started to attract attention indicating
emergence of quantum plasmonics as new field of research.\cite{StockmanOE11,TameNatPhys13,NordlanderNatMater10,AbajoNature12,SchollNature12,AizpuruaBaumbergNature12,NordlanderScience14,ElKhouriNL14,BaiBosmanNijhuisScience14,AizpuruaCrozierNatComm16,AizpuruaBaumbergScience16,ReinhardAdvMat16,NijhuisAppMatToday16}
Similarly, strong light-molecule coupling in nanocavities,\cite{SchwartzEbbesenPRL11,SchwartzEbbesenAngChemIntEd12,SchwartzEbbesenAdvMat13,EbbesenChemPhysChem13,EbbesenAngChemIntEd13} 
when states of light and matter cannot be 
separately distinguished and a hybrid state (polariton) is formed,
reveals quantum nature of external electromagnetic field.
Finally, very recently ultra-strong coupling regime (the regime where the coupling between
light and matter becomes the largest energy scale in the system) was achieved
experimentally.\cite{MurchNatPhys17} 

Experimental advancements in nanojunction spectroscopy posed a challenge for adequate theoretical 
description.\cite{RatnerMRSBulletin04,MGANPCCP12,SchatzRatnerRepProgPhys12}
In particular, these advancements resulted in necessity to combine theoretical tools of optical 
spectroscopy with those of quantum transport.
Corresponding formulations were developed and applied to description of absorption
and current-induced light emission,\cite{GalperinNitzanPRL05,GalperinNitzanJCP06,Harbola2006b,GalperinTretiakJCP08,MiwaJPhysSocJpn13,MiwaJPhysSocJpn13_2,MiwaNanoResLett13,MiwaEJSurfSciNano15,DubiSciRep15,MukamelHarbolaJCTC15,DongChinJChemPhys15,NiehausJPCC16,MiwaNature16,YamNanoscale16,PavanJPCC16}
as well as light-induced current in junctions.\cite{GalperinNitzanPRL05,FainbergNitzanPRB07,KleinekathoferPRB08,FainbergSukharevParkMGPRB11,PeskinMGJCP12,WhitePeskinMGPRB13,SelzerPeskinJPCC13,OchoaSelzerPeskinMGJPCL15,RabitzJCP14,RatnerSeidemanJCP14,VenkataramanPCCP16}
In these studies light-matter interaction mostly was treated combining classical electrodynamics 
of radiation field with quantum mechanical description of the molecule.\cite{NeuhauserJCP09,LopataNeuhauserJCP09,SukharevMGPRB10,RatnerSchatzJPCC10,NeuhauserJCP11,SukharevNitzanPRA11,WhiteSukharevMGPRB12,GaoNeuhauserJCP12,SchatzRatnerRepProgPhys12,SchatzJPCA12,NitzanJCP13,RatnerSeidemanJCP14,RatnerSeidemanNL14,DongNanoscale15}
Similarly, theoretical approaches to (yet to be measured in junctions) multidimensional
spectroscopy were proposed.\cite{RahavMukamelJCP10,MukamelJCP15,MukamelJCP16,BittnerSciRep16,BittnerChemPhys16}

Significant theoretical efforts were devoted to development of theory of Raman spectroscopy
in current-carrying junctions\cite{GalperinRatnerNitzanNL09,GalperinRatnerNitzanJCP09,HarbolaMukamelJCP14,MGRatnerNitzanJCP15,GaoMGANJCP16,ApkarianMGANPRB16}
and its application to modeling of current induced heating,\cite{GalperinRatnerNitzanNL09,GalperinRatnerNitzanJCP09,MGANJPCL11,MGANPRB11,WhiteTretiakNL14}
dynamics and conformational changes,\cite{BrandbygeHedegardTodorovDundasPRB12,YinJRS12,BaumbergNL15}
chemistry,\cite{WuChemComm11}
control of charging states of the molecule,\cite{RatnerJPCC12,WhiteTretiakNL14,KronikNeatonNatelsonPNAS14,KronikNeatonNatelsonNL16}
study of time-dependent correlations between conductance and Raman,\cite{ParkMGEPL11,ParkMGPRB11,ParkMGPST12}
and elucidation of chemical enhancement in SERS.\cite{JensenSchatzNL06, OrenMGANPRB12,ParkMGEPL11,ParkMGPRB11,BanikApkarianParkMGJPCL13,ApkarianJPCC12,VanDuyneSchatzJPCL13,MujicaJPCC14}
Here radiation field was mostly treated quantum mechanically.
Also strong light-matter (plasmon-molecule) couplings were treated theoretically
with radiation field described quantum mechanically.\cite{NordlanderNL09,SavastaPRL10,NordlanderNL11,NordlanderJCP11,NordlanderNatCommun12,NordlanderNL12,WhiteFainbergMGJPCL12,NordlanderPRB12,ZabalaOptExpress13,MukamelNatComm13,LaussyPRA16_1,LaussyPRA16_2,MukamelJCP16,KowalewskiMukamelJCP16}.
Finally, quantum treatment of the field was required to describe optically measured 
noise characteristics of junctions.\cite{BelzigPRL14,KaasbjergNitzanPRL15}
Quantum effects in photonics and optical spectroscopy were discussed in recent reviews.\cite{LodahlRMP15,MukamelRMP16}

From theoretical perspective optical spectroscopy in nanojunctions (optoelectronics)
is a field where theoretical approaches of nonlinear optical spectroscopy meet those 
of quantum transport theory. Theoretical toolboxes of the two research communities 
are slightly different and sometimes also utilize a bit different language. For example, 
traditionally optical spectroscopy relies on bare perturbation theory (PT) in the Liouville 
space for classification of optical response of isolated molecular systems.
Transitions in the latter are considered in the basis of many-body states of the molecule. 
This type of treatment became standard in the spectroscopy community.
Naturally, this same approach is sometimes applied to open systems and/or 
when radiation field is treated quantum mechanically.
Theoretical methods of quantum transport community are numerous.
The common (and probably most developed) is the nonequilibrium Green's functions (NEGF) approach.
It is formulated in the  Hilbert space and in its canonical form utilizes 
quasiparticles (or elementary excitations, or molecular orbitals) as a basis.

This review compares theoretical approaches of the two communities in
their treatment of spectroscopy in nanojunctions. We discuss their strong and weak sides  
and indicate limitations in applicability of the approaches.
Structure of the review is the following. 
Section~\ref{spectroscopy} discusses theoretical approaches to optical spectroscopy.
To make the review self-contained in Section~\ref{liouville} we give
a short introduction to theoretical methodology standard in the nonlinear optical spectroscopy
community. Section~\ref{hilbert} discusses photonics in nanojunctions from perspective
of Green function methods. We first shortly introduce canonical NEGF in Section~\ref{negf}
and then follow with its two many-body flavors: the pseudoparticle NEGF (PP-NEGF)
in Section~\ref{ppnegf} and Hubbard NEGF in Section~\ref{hubnegf}.
We compare the theoretical approaches and argue that the latter two formulations
can be a convenient choice for optoelectonics.
Theoretical considerations of spectroscopy in junctions
with radiation field treated classically are presented in  Section~\ref{classical}.
Quantum treatment of radiation field in junctions is discussed in Section~\ref{quantum}.
Section~\ref{conclude} concludes.

\section{Theoretical methodology}\label{spectroscopy}
A distinct feature of junction spectroscopy is mixing between 
optical and electronic characteristics of an optoelectronic device. 
Indeed, optical spectroscopy of isolated systems is focused on photon flux; this flux is the only
channel of communication between the system and environments (measuring devices). 
On the contrary, in nanojunctions, where electron participating in optical scattering
process is free to leave contributing to electron and energy fluxes, 
combined theoretical consideration of all the constituents is crucial. 
In this case one does not have optical signal independent of electric current, rather
one has to deal with a comprehensive description. 

Let consider junction under external illumination.\footnote{For future reference we write down a quantized radiation field.} 
Hamiltonian of the total system is
\begin{equation}
\label{H}
\hat H=\hat H_M + \hat H_K + \hat H_p + \hat V_{MK} + \hat V_{MP} 
\end{equation}
Here $\hat H_M$, $\hat H_K$, and $\hat H_p$ are, respectively, matter (e.g., molecular),
contacts, and radiation field Hamiltonians. $\hat V_{MK}$ and $\hat V_{MP}$ describe
coupling to contacts and light-matter interaction. In general part or all of the contributions
can be time-dependent due to external driving. Contacts and radiation field Hamiltonians are 
assumed to be reservoirs of free carriers (electrons and photons, respectively):
$\hat H_K=\sum_{k\in K}\varepsilon_k\hat c_k^\dagger\hat c_k$ and
$\hat H_p=\sum_\alpha \hbar\omega_\alpha\hat a_\alpha^\dagger\hat a_\alpha$,
where $\hat a_\alpha^\dagger$ ($\hat a_\alpha$) and $\hat c_k^\dagger$ ($\hat c_k$)
are creation (annihilation) operators for photon in mode $\alpha$ and electron in state $k$
of contact $K$. 
For simplicity below we specialize to bilinear coupling 
coupling to contacts and rotating wave approximation in light-matter interaction\footnote{Note that more general couplings can be considered as well.\cite{GalperinNitzanPRL05,FainbergSukharevParkMGPRB11}} 
\begin{align}
\label{VMP}
\hat V_{MP}=&\sum_{m\in M}\sum_{\alpha}(U_{m\alpha}\hat X_m^\dagger\,\hat a_\alpha + H.c.)
\\
\label{VMK}
\hat V_{MK}=&\sum_{m\in M}\sum_{k\in K}(V_{mk}\hat X_m^\dagger\,\hat c_k+H.c.), 
\end{align}
where $\hat X_m^\dagger$ are matter excitation operators 
due to electron transfer from state $k$ in contacts or optical excitation
by radiation field. 

Most theoretical studies in optical spectroscopy and quantum transport are focused on
evaluation of fluxes (photon, $p$, and electron, $e$, respectively), which are
defined as rates of change of carriers populations in baths (respectively,
photon population of radiation field modes and electron density in contacts)\cite{Mukamel_1995,HaugJauho_2008}
\begin{align}
\label{Ip}
 I_p(t) =& -\frac{d}{dt}\sum_\alpha \langle \hat a_\alpha^\dagger(t)\hat a_\alpha(t)\rangle
 =2\,\mbox{Im}\sum_{m\in M}\sum_{\alpha} U_{m\alpha}\langle\hat X_m^\dagger(t)\hat a_\alpha(t)\rangle
 \\
 \label{IK}
 I_e^K(t)=& -\frac{d}{dt}\sum_{k\in K}\langle \hat c_k^\dagger(t)\hat c_k(t)\rangle
 =2\,\mbox{Im}\sum_{m\in M}\sum_{k\in K} V_{mk}\langle\hat X_m^\dagger(t)\,\hat c_k(t)\rangle
\end{align} 
Here , the operators are in Heisenberg picture and 
$
\langle\ldots\rangle = \mbox{Tr}[\ldots\hat \rho(t_0)]\ \
$
 is quantum mechanical and statistical averaging with respect to initial density operator 
(usually assumed to be direct product of radiation field and electronic components
$\hat\rho(t_0)=\hat \rho_p(t_0)\otimes\hat \rho_e(t_0)$).
Energy fluxes, $J_p(t)$ and $J_e^K(t)$, are defined in a similar way as rates of change of energy in the baths.\footnote{However, note recent discussion on inconsistency of this definition with thermodynamic laws.\cite{EspOchoaMGPRB15,NitzanPRB16}}
The fluxes can be expressed in terms of single-particle Green functions (two time correlation
functions; see below).
Note that index $m$ in Eqs.~(\ref{VMP}) and (\ref{Ip}) has a meaning of {\em optical transfer}
in the system, that is total number of electrons in $M$ does not change. On the contrary,
$m$ in Eqs.~(\ref{VMK}) and (\ref{IK}) indicates {\em electron transfer} between $M$ and $K$;
such transfer results in change of electron population in the system. 

Other quantities of interest are related to statistics of photon\cite{GlauberPRL63,GlauberPhysRev63,KleinerPR64,ScullyLambPR69} 
and electron\cite{Levitov1993,Levitov1996,BlanterButtikerPhysRep00} transport as well as
cross-correlations between the two.
Measurements of fluctuations of particle fluxes were reported in 
junction studies for photon\cite{YamamotoNature99,PortierPRL10} and electron transport.\cite{RuitenbeekNL06,TalPRL08,ForguesSciRep13,BerndtSurfSci16}
Number of experiments demonstrated cross-correlation effects.\cite{FujisawaScience06,BerndtPRL10,BerndtPRL12}
Theoretically fluctuations are characterized within the full counting statistics 
(for time-local cumulants of transfer distribution)\cite{EspositoRMP09} 
or via two-particle Green function,
(e.g., $g(2)$ \cite{BrandesPRB12,AnkerholdPRB15} or 
current-current\cite{BoGalperinJPCM96,ImryPRB00,GavishImryLevinsonARXIV02,SouzaJauhoEguesPRB08} 
correlation functions).
Higher order correlation functions were also considered in the literature.\cite{DorfmanMukamelPhysScr16} 

Evaluation of the correlation functions is performed within either Liouville or Hilbert spaces
with the former being standard choice in the nonlinear spectroscopy community.
While the two representations are differ only in the way correlation function is evaluated, 
and ideally one expects the same result from both considerations, approximations
involved in real-life calculations are quite different, so that results depend on the way (and level) 
of treatment. Below we give a short pedagogical introduction to different approaches
and indicate their strong and weak sides.   

\subsection{Liouville space formulation}\label{liouville}
We start from a very short introduction to the Liouville space formulation - an accepted standard 
approach in the field of nonlinear optical spectroscopy. A comprehensive formulation
can be found in Ref.~\cite{Mukamel_1995}, which became a standard reference 
for classification and interpretation of optical experiments. 

\begin{figure}[htbp]
\centering
  \includegraphics[width=\linewidth]{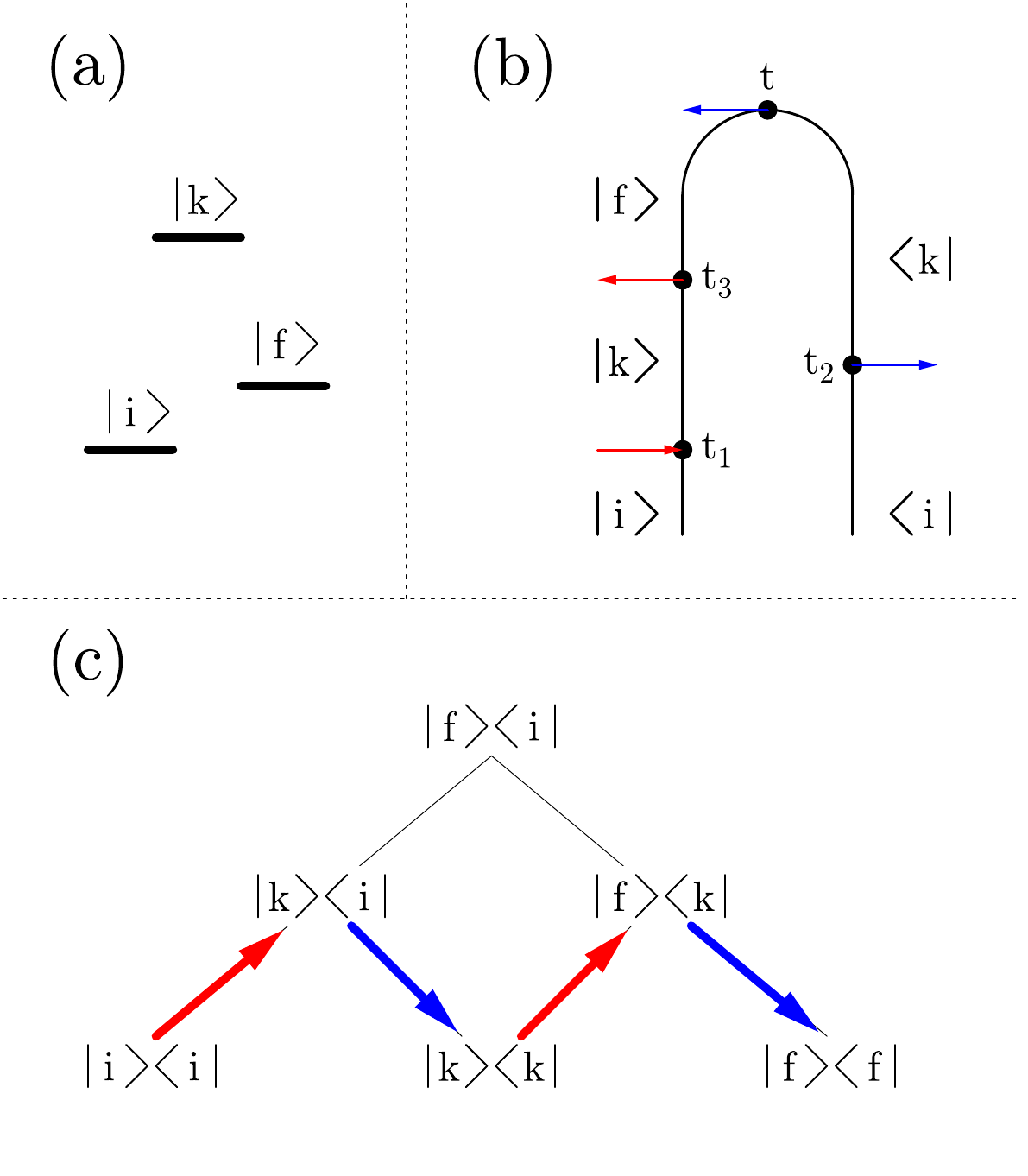}
  \caption{Liouville space formulation of an optical spectroscopy. Shown are 
  (a) three level system $\hat H_M$, (b) double-sided Feynman diagram representing 
  a spontaneous light emission process, and (c) corresponding Liouville space pathway.}
  \label{fig1}
\end{figure}

Expressing photon flux, Eq.(\ref{Ip}), in interaction picture with respect to the light-matter coupling (\ref{VMP}) yields $I_p(t)=2\,\mbox{Im} \sum_{m,\alpha} U_{m\alpha} \mbox{Tr}[\hat X^\dagger_{I,m}(t)\,\hat a_{I,\alpha}(t)\,\hat \rho_I(t)]$, where subscript $I$ indicates interaction picture.
Integral form of the Liouville-von Neumann equation for the total density matrix is
\begin{align}
\label{rhoI}
\begin{split}
\lvert \rho_I(t)\rangle\rangle=&T\,\exp\left[-i\int_{t_0}^tds\,\mathcal{V}_{I,MP}(s)\right]
\lvert\rho(t_0)\rangle\rangle
\\
\equiv& \left[1+\sum_{n=1}^{\infty} (-i)^n\int_{t_0}^{t} dt_n\int_{t_0}^{t_n}dt_{n-1}\ldots\int_{t_0}^{t_2}dt_1
\right.
\\ & \left.
\mathcal{V}_{I,MP}(t_n)\mathcal{V}_{I,MP}(t_{n-1})\ldots\mathcal{V}_{I,MP}(t_1)\right]\lvert\rho(t_0)\rangle\rangle
\end{split}
\end{align}
Here $T$ is the time ordering operator, expression is written in the Liouville space,
and $\mathcal{V}_{I,MP}$ is interaction picture form of the superoperator corresponding to 
the Hilbert space operator $\hat V_{MP}$ of (\ref{VMP}). 
Expansion of the evolution operator ($T$ ordered exponent in first row of Eq.(\ref{rhoI})) 
in Taylor series (second and third rows of (\ref{rhoI}))
yields bare perturbation theory (PT) in $\hat V_{MP}$, which presents evolution of
the total density matrix as a sum of time-ordered ($t\geq t_n\geq t_{n-1}\geq\ldots \geq t_1\geq t_0$) sets of light-matter interactions. Taking into account that any Liouville space superoperator
is expressed in Hilbert space as a commutator of the corresponding operator,
every $\mathcal{V}_{I,MP}(t_i)$ in (\ref{rhoI}) acts on either bra or ket of the result of prior evoluton,
$\int_{t_0}^{t_i} dt_{i-1}\ldots\int_{t_0}^{t_2} dt_1\mathcal{V}_{I,MP}(t_{i-1})\ldots \mathcal{V}_{I,MP}(t_1)\lvert\rho(t_0)\rangle\rangle$. 
Thus, Liouville space expression for photon flux (\ref{Ip}),
\begin{equation}
\label{IpLiouville}
 I_p(t)=2\,\mbox{Im}\sum_{m\in M}\sum_\alpha \langle\langle a_{I,\alpha}^\dagger(t) X_{I,m}(t)\vert
 \rho_I(t)\rangle\rangle
 \equiv \sum_{n=0}^\infty I_p^{(n)}(t),
\end{equation}
can be conveniently persented on the Schwinger-Keldysh contour\cite{SchwingerJMP61,KeldyshSovPhysJETP65}
as sets of points representing interaction with radiation field at times $t_1$,\ldots, $t_n$
on forward (bra evolution) and backward (ket evolution) branches (see Fig.~\ref{fig1}). 
Time of observation $t$ naturally belongs to both branches.
By convention arrows indicate creation ($\hat a_\alpha^\dagger$, pointing to left) 
or annihilation ($\hat a_\alpha$, pointing to right) photon operators for quantum radiation field, 
or corresponding negative and positive components when radiation field is treated classically.
Graphical representations of the type shown in Fig.~\ref{fig1}b,
coined double-sided Feynman diagrams,\footnote{Note that the name is a bit misleading, because for quantum radiation field a diagram is characterized also by contractions (Green function) of corresponding field operators (see Fig.~\ref{fig3} below).} 
are widely utilized in the spectroscopy community to  classify optical processes.
Time ordered sequences of changes in the state of the system (changes in its bra and ket with time)
are known as pathways in Liouville space (see Fig.~\ref{fig1}c); they are instrumental in discussing
propagation of coherences and populations in the system resulting from optical scattering processes. 
Similarly, one can expand electron flux $I_e^K(t)$, Eq.~(\ref{IK}), in orders of light-matter interaction
$\hat V_{MP}$, Eq.(\ref{VMP}). Expansions in coupling to contacts $\hat V_{MK}$, 
Eq.(\ref{VMK}), were also considered in the literature.\cite{MukamelHarbolaJCTC15}  

Bare PT (\ref{rhoI}) in $\hat V_{MP}$ decouples light and matter degrees of freedom,
i.e. each contribution is a product of two correlation functions (electron and photon), 
which have to be evaluated independently. Often after completing the derivation
incoming field is assumed to be in a coherent state, and transfer to classical representation
is performed. Thus, optical response to classical field only requires 
evaluation of the electronic multi-time correlation function.
Even contributions in the expansion (\ref{IpLiouville}) usually drop out because
of the odd number of photon creation/annihilation operators in the correlation function
(for classical field these terms drop out by symmetry in the case of isotropic medium\cite{Mukamel_1995}).
Sometimes quantum description is used for a subset of modes, while the rest of the field 
is treated classically.\cite{DorfmanSchlawinMukamelJPCL14}
Note that application of bare PT to description of quantum fields 
(or to molecule-contacts coupling $\hat V_{MP}$) may be problematic
even when perturbation theory is applicable (e.g., when ratio of light-matter coupling is small
compared with the system coupling to electronic bath). The reason is ability of photons
to serve as an intermediates inducing effective non-Markov interactions within electron system.
The latter enters theoretical description via electron self-energies, which cannot be properly
described within bare PT (see discussion in Section~\ref{negf}).

Contrary to isolated systems, in junctions electron correlation is averaged over both
system (molecule) and bath (contacts) degrees of freedom. As long as light-matter interaction 
is assumed to be  confined to the system (molecule) only (i.e. electron operators in the correlation 
function are those of the system only), the multi-time correlation function can be evaluated by 
employing the regression formula. This procedure is often utilized and is exact when system evolution 
is Markovian.\cite{BreuerPetruccione_2003}
Note that while formally time-local quantum master equation can always be derived, in practice
evaluation of the time-convolutionless propagator is a complicated task.\cite{RabaniJCP15}
Thus, most practical applications so far utilize effective Markovian propagators,
which (when employed with the regression formula) may be problematic.
In particular,  within the approach every interaction with the optical field results in the destruction of 
the molecule-contacts coherence. The latter is an artifact of the formulation, which 
may lead to qualitative failures (see discussion in Section~\ref{classical}). 


To summarize, 
the Liouville space superoperator formulation of nonlinear optical spectroscopy has several
important advantages. First, the formulation follows evolution of the density matrix in real time,
which allows for an intuitive graphical representation of optical processes in the form of 
double sided Feynman diagrams. Second, system's response can be described within
the basis of its many-body states (eigenbasis of the Hamiltonian $\hat H_M$), which
allows to account for all the intra-system interactions exactly and paves a way to incorporate
results of standard quantum-chemistry simulations (usually performed for isolated molecules)
into numerical modelling of optoelectronic devices.
These advantages were mentioned in many works on optical spectroscopy.\cite{Mukamel_1995,MukamelPRE03,Harbola2006b,HarbolaMukamelPhysRep08,HarbolaMukamelJCP14,MukamelJCP15,MukamelHarbolaJCTC15,MukamelJCP16}
At the same time the bare PT and utilization of the regression formula may be problematic.
The former fails to adequately describe open systems when radiation field is treated quantum mechanically.
The latter leads to qualitative mistakes due to approximate nature of effective Markovian propagators
employed in practical simulations. 
This is true for both classical and quantum radiation field treatments.
Note that multi-time correlation functions in principle can be
evaluated numerically exactly (i.e. without employing the regression formula).\cite{RabaniPRL08,WegewijsPRB12,WegewijsPRB14,CohenGullReichmanMillisPRL15}
However,  significant cost of such approaches so far limits their applicability to simple models.
Also, standard approach (e.g., Redfield quantum master equation), which employs 
many-body states of the system as a basis, does not 
properly account for the system (molecule) - electronic baths (contacts) couplings.\cite{EspGalpPRB09,EspositoMGJPCC10}

Below we show that Hilbert space Green function formulations are capable to yield
(within similar level of theory)
the same advantages while avoiding pitfalls of the Liouville space superoperator methods.   

\subsection{Hilbert space formulations}\label{hilbert}
While theoretical treatments of quantum transport utilize both Hilbert and Liouville space formulations,
the former is the choice of the nonequilibrium Green function (NEGF) technique. Below we focus on
the NEGF (and its generalizations) and discuss application of the technique to description of
optical spectroscopy in nanojunctions. 

\begin{figure}[htbp]
\centering
  \includegraphics[width=\linewidth]{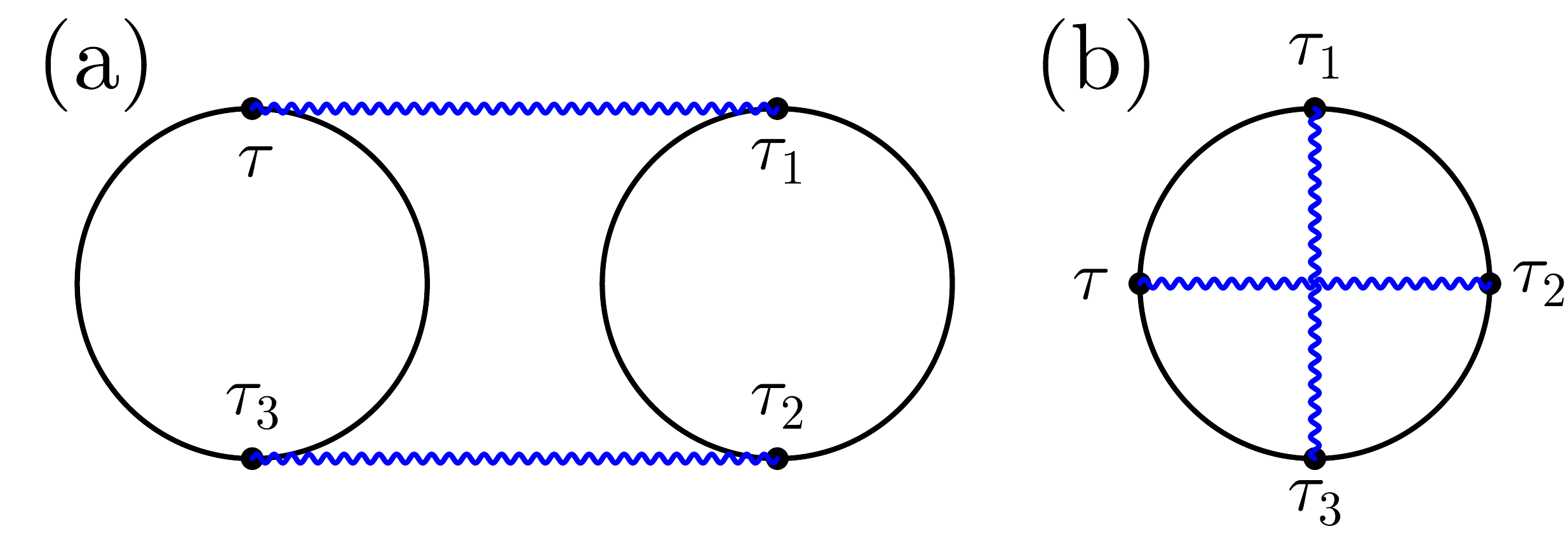}
  \caption{Feynman diagrams for (a) two-particle scattering and (b) virtual photon exchange.
  Solid (black) and wavy (blue) lines represent electron and photon Green function, respectively.}
  \label{fig2}
\end{figure}

Evolution of a nonequilibrium system in the Hilbert space
relies on contour (rather than real time as in Liouville space) ordering. 
A particular ordering of contour variables, which we will denote by Greek index $\tau$ in contrast to $t$
indicating real time, defines projection of a correlation function. These projections are equivalent
to double sided Feynman diagrams of the Liouville space formulation, although number of 
contour projections is smaller (requirement of real time ordering in the Liouville space formulation 
results in bigger number of such projections, i.e. one contour projection includes several
double sided Feynman diagrams). Note that projections and diagrams (in their original meaning)
are different things. For example, Fig.~\ref{fig2} demonstrates two different fourth order Feynman 
diagrams representing two-particle scattering and virtual photon exchange processes, respectively; 
both diagrams will have the same set of projections.

Historically introduction of the contour is a consequence of an attempt to build analog of 
Feynman diagrams for nonequilibrium systems.
Feynman diagrammatic technique relies on the Gell-Mann and Low theorem\cite{FetterWalecka_1971} 
which (for equilibrium  system at zero temperature) allows to establish connection between 
initial ($t_0\to-\infty$) and final ($t_0\to+\infty$) ground states of the system. 
Absence of such connection in nonequilibrium 
situation
necessitates using bra and ket of the starting state 
as initial and final states of the process (i.e. starting state density matrix is utilized for quantum mechanical 
and statistical averaging).\footnote{Note that if initial correlations are to be taken into account 
one needs to consider a combination of the Keldysh and Matsubara contours. 
We do not discuss this possibility here, an interested reader is encouraged to consult Refs.~\cite{Danielewicz1984,WagnerPRB91,vanLeeuwenJCP09,LeeuwenJPCS10}.}
Ref.~\cite{WagnerPRB91} presents a beautiful and thorough 
discussion about relations between the Feynman (zero temperature equilibrium), 
Matsubara (finite temperature equilibrium),  and Keldysh (nonequilibirum) theories.

Fluxes (\ref{Ip}) and (\ref{IK}) can be {\em exactly} expressed in terms of Green functions as\cite{MeirWingreenPRL92,JauhoWingreenMeirPRB94,HaugJauho_2008,GaoMGJCP16_1}
\begin{align}
\label{IpGF}
I_p(t) =& -2\,\mbox{Re}\sum_{\alpha_1,\alpha_2}\int_{t_0}^t ds\,
\left[ \Pi^{<}_{\alpha_1\alpha_2}(t,s)\, F_{\alpha_2\alpha_1}^{>}(s,t) 
\right. \\ & \left. \qquad\qquad\qquad\qquad -
        \Pi^{>}_{\alpha_1\alpha_2}(t,s)\, F_{\alpha_2\alpha_1}^{<}(s,t) \right]
\nonumber \\
\label{IKGF}
I^K_e(t) =& 2\,\mbox{Re}\sum_{m_1,m_2\in M}\int_{t_0}^t ds\,
\left[ \Sigma^{K,<}_{m_1m_2}(t,s)\, G_{m_2m_1}^{>}(s,t)
\right. \\ & \left. \qquad\qquad\qquad\quad -
        \Sigma^{K,>}_{m_1m_2}(t,s)\, G_{m_2m_1}^{<}(s,t) \right]
\nonumber
\end{align}
where $<$ ($>$) are lesser (greater) projections of electron $G$ and photon $F$ Green functions 
as well as electronic self-energy due to coupling to contact $K$, $\Sigma^K$, 
and photon self-energy due to coupling to electrons, $\Pi$.
Their explicit on-the-contour definitions are
\begin{align}
 \label{defG}
 &G_{m_1m_2}(\tau_1,\tau_2) = -i\left\langle T_c\, \hat X_{m_1}(\tau_1)\,\hat X_{m_2}^\dagger(\tau_2)\right\rangle
 \\
 \label{defF}
 &F_{\alpha_1\alpha_2}(\tau_1,\tau_2) = -i\left\langle T_c\, \hat a_{\alpha_1}(\tau_1)\,\hat a_{\alpha_2}^\dagger(\tau_2)\right\rangle
 \\
 \label{defSK}
 &\Sigma^{K}_{m_1m_2}(\tau_1,\tau_2) = \sum_{k\in K} V_{m_1k}\,g_k(\tau_1,\tau_2)\, V_{km_2}
 \end{align}
Here $T_c$ is the contour ordering operator, $\tau_{1,2}$ are contour variables, and
$g_k(\tau_1,\tau_2)\equiv- i\langle T_c\,\hat c_k(\tau_1)\,\hat c_k^\dagger(\tau_2)\rangle$
is the Green function of free electron in state $k$ of contact $K$.
Explicit expression for the photon self-energy, $\Pi_{\alpha_1,\alpha_2}(\tau_1,\tau_2)$, 
depends on the level of treatment (e.g., order of diagrammatic perturbation theory employed - see below).
While analytical forms of the self-energy $\Sigma^K$ is known, other constituents of 
the fluxes expressions (\ref{IpGF}) and (\ref{IKGF}) have to be evaluated by solving a
set of coupled equations.

Note that expression for photon flux, Eq.(\ref{IpGF}), is relevant only when radiation field is treated 
quantum mechanically. For classical field one either has to evaluate multi-time correlation functions
as discussed in section~\ref{liouville}, or solve time-dependent problem for a system (molecule)
coupled to external classical field. The latter situation was considered in many works,\cite{NeuhauserJCP09,LopataNeuhauserJCP09,SukharevMGPRB10,NeuhauserJCP11,FainbergSukharevParkMGPRB11,SukharevNitzanPRA11,GaoNeuhauserJCP12,WhiteSukharevMGPRB12}
where radiation field was treated classically propagating Maxwell equations along with
quantum mechanical treatment of electron dynamics.
Expression for electron flux, Eq.(\ref{IKGF}), is always correct.
Note also that in junctions where formation of the local field is affected by both
plasmon excitations in the contacts and molecular response 
bare perturbation theory (which for a particular optical process
discards back action of the matter on the field) may be not enough.
This was shown in studies considering radiation field both classically\cite{NeuhauserJCP11,WhiteSukharevMGPRB12} and
quantum mechanically.\cite{WhiteFainbergMGJPCL12}
Finally, the fact that, e.g., electron Green functions $G$ enters expressions for both fluxes,\footnote{Electron Green function enters expression for photon flux, Eq.(\ref{IpGF}), via photon self-energy $\Pi$.}
Eqs.~(\ref{IpGF}) and (\ref{IKGF}), indicates inter-dependence of the fluxes
and demonstrates a necessity of consistent (i.e. within the same
level of theory) description of  an optoelectronic device responses. 

\subsubsection{Nonequilibrium Green functions (NEGF)}\label{negf}
NEGF utilizes language of quasipartcles  (or elementary excitations, or orbitals):
$d^\dagger_i$ ($\hat d_i$) are the usual operators of the second quantization 
which create (annihilate) electron in orbital $i$ of the system $M$.  
The approach is most convenient when intra-system interactions are small compared with
the coupling to electronic baths, so that the former can be treated within diagrammatic 
perturbation theory. In this case one can utilize, e.g., set of molecular (or Kohn-Sham) 
orbitals to represent electronic structure of the molecule. The latter are (de)populated by 
electron transfer between electronic reservoirs and molecule. 
Thus index $m$ in Eq.(\ref{IKGF}) stands for such an orbital, 
so that electron flux is exactly expressed in terms of single particle Green functions
\begin{equation}
 \label{defGNEGF}
 G_{ij}(\tau_1,\tau_2) = -i\langle T_c\,\hat d_i(\tau_1)\,\hat d_j^\dagger(\tau_2)\rangle
\end{equation}
Green functions (\ref{defF}) and (\ref{defGNEGF}) are defined by
solving Dyson equations
\begin{align}
\label{GDyson}
 &G_{ij}(\tau_1,\tau_2)= G_{ij}^{0}(\tau_1,\tau_2)
 \\ & \quad
 +\sum_{k,m}\int_c d\tau_3\int_c d\tau_4\,
 G_{ik}^{0}(\tau_1,\tau_3)\,\Sigma^p_{km}(\tau_3,\tau_4)\,G_{mj}(\tau_4,\tau_2)
\nonumber  \\
\label{FDyson}
 &F_{\alpha_1,\alpha_2}(\tau_1,\tau_2) = F_{\alpha_1,\alpha_2}^{0}(\tau_1,\tau_2)
 \\ & \quad
 +\sum_{\alpha_3,\alpha_4}\int_c d\tau_3\int_c d\tau_4\,
 F_{\alpha_1,\alpha_3}^{0}(\tau_1,\tau_3)\,\Pi_{\alpha_3\alpha_4}(\tau_3,\tau_4)\,
 F_{\alpha_4,\alpha_2}(\tau_4,\tau_2)
 \nonumber
 \end{align}
 Here $G^0$ and $F^0$ are the Green functions in the absence of the light-matter coupling $\hat V_{MP}$,
 $\Sigma^p$ is electronic self-energy due to coupling to radiation field,
 and $\Pi$ is photon self-energy due to coupling to electrons. 
The two latter quantities can be derived only approximately; the approximations should 
satisfy conservation laws for physical quantities. A way to formulate conserving approximations
was formulated in the works by Kadanoff and Baym.\cite{BaymKadanoffPR61,BaymPR62}
An interested reader is encouraged to consult book~\cite{StefanucciVanLeeuwen_2013}
for a detailed consideration of the issue.

Diagrammatic perturbation theory is built by constructing the Luttinger-Ward functional, $\Phi$.
The latter is collection of all dressed connected skeleton diagrams (i.e., connected combinations of 
Green functions $G$ and $F$ that have no self-energy insertions).\cite{LuttingerWardPR60,DeDominicisMartinJMP64}
Expressions for self-energies are obtained as functional derivatives\cite{Haussmann_1999,StefanucciVanLeeuwen_2013}
\begin{align}
\label{SpPhi}
\Sigma^p_{m_1m_2}(\tau_1,\tau2) =& 
\frac{\delta\Phi[G,F]}{\delta G_{m_2m_1}(\tau_2,\tau_1)}
\\
\label{PiPhi}
\Pi_{\alpha_1,\alpha_2}(\tau_1,\tau_2) =& 
-\frac{\delta\Phi[G,F]}{\delta F_{\alpha_2\alpha_1}(\tau_2,\tau_1)}
\end{align}
An example of fourth order (in light-matter interaction) contribution to the functional $\Phi$  
is shown in Fig.~\ref{fig2}b. Functional derivatives in Eqs.~(\ref{SpPhi}) and (\ref{PiPhi})
corresponds to removal of one straight (Green function $G$) or wavy (Green function $F$)
line, respectively. 

It is important to note that contrary to bare PT (an expansion to a particular order), 
diagrammatic PT sums a particular type of diagrams (the type corresponding to a particular
order) to infinity.
Indeed, Luttinger-Ward functional $\Phi$ is expressed in terms of full Green functions $G$ and $F$ 
rather than their zero order analogs  $G^0$ and $F^0$.\footnote{Note  that also structure of the Dyson equations,  Eqs.~(\ref{GDyson}) and (\ref{FDyson}), implies resumming which accounts for reducible diagrams.} 
This summation is central for conserving character of an approximation. 
Note also that necessity (for quantum radiation field) to solve simultaneously two coupled 
(via their self-energies) Dyson equations (\ref{GDyson}) and (\ref{FDyson}) corresponds
to formation of the local field due to both external source and system response.
The situation is relevant for, e.g., molecules in nanocavities or in vicinity of metal surfaces.
However, even if one can assume independence of the radiation field on molecular response\footnote{One has to use $F^0$ everywhere in this case and disregard Eq.(\ref{FDyson}).},
still for an approximation to be conserving one has to sum a subset of diagrams to infinity -
via dependence of electron self-energy $\Sigma^p$ on full Green function $G$ and
structure of the Dyson equation (\ref{GDyson}). Bare PT does not take into account these resummations 
and thus will violate conservation laws. As a result traditional classification of optical processes
(and utilization of the double sided Feynman diagrams as is standard in optical spectroscopy community) 
becomes questionable in comprehensive treatments of open systems. 


To summarize, an important advantage of the NEGF is existence of established set of rules 
(the nonequilibrium diagrammatic technique) to treat  interactions in the system 
(in particular, light-matter interaction) in an organized  perturbative way preserving
physical conservation laws. The approach is capable to exactly account for system (molecule) coupling 
to electronic baths (contacts). It is also instrumental in studying counting statistics of transport
both in steady-state\cite{GogolinKomnikPRB06} and transient\cite{TangWangPRB14,TangWangPRB16} 
regimes. NEGF is direct successor to original (Feynman and Matsubara) Green function
formulations\cite{WagnerPRB91}  and as such allows physically motivated choice of relevant diagrams 
in the same way as does the Feynman diagrammatic technique. 
The main drawback of the method with respect to optoelectronic problems is its qausiparticle
formulation. This significantly complicates treatment of strong intra-system interactions
and makes (TD)DFT (also formulated in the basis of effective single particle orbitals) 
a method of choice for electronic structure simulations.\cite{DamleGhoshDattaCP02,XueDattaRatnerCP02,TaylorStokbroPRB02}
The latter does not connect well with traditional optical spectroscopy language
and has some  limitations related to both foundations of (TD)DFT 
(e.g., utilization of Kohn-Sham orbitals as physical objects) and in terms of
its applications to transport.\cite{BaratzMGBaerJPCC13} 

We now turn to many-body flavors of the NEGF which to some extent are capable of overcoming 
these limitations.  

\subsubsection{Pseudoparticle NEGF}\label{ppnegf}
Contrary to creation (annihilation) of quasiparticles in orbitals
in the usual second quantization, pseudoparticle operators create (annihilate)
many-body states of a system (molecule). For example, pseudoparticle S (corresponding
to eigenstate $\lvert S\rangle$ of the Hamiltonian $\hat H_M$) is
constructed by applying creation operator $\hat p^\dagger_S$ to unphysical vacuum $\lvert vac\rangle$.
Pseudoparticle operators satisfy the usual commutation or anticommutation relations depending 
on the Bose (e.g., even number of electrons) or Fermi (e.g., odd number of electrons) character 
of the corresponding many-body state. Any full set of many-body states
should fulfill normalization condition (sum over many-body states probabilities should be one).
This condition is not automatically satisfied by the second quantization in the space of many-body
states of the system, and should be imposed to restrict the so called extended Hilbert space to
its physical subspace
\begin{equation}
 \label{Q1}
 \hat Q\equiv \sum_S \hat p_S^\dagger\hat p_S=1
\end{equation}
Naturally, pseudoparticle representation diagonalizes the system Hamiltonian, 
$\hat H_M=\sum_S E_S\hat p_S^\dagger\hat p_S$.
However, it makes $\hat V_{MK}$ non-quadratic: index $m$  in Eq.~(\ref{VMK})
indicates transition $(S_1,S_2)$ between pair of states which differ by single electron,
so the interaction becomes $\hat V_{MK}=\sum_{S_1,S_2\in M}\sum_{k\in K}\left(V_{(S_1S_2),k}\hat p^\dagger_{S_2}\hat p_{S_1}\hat c_k+H.c.\right)$.
Similarly, index $m$ in Eq.~(\ref{VMP}) stands for a transition between pair of states with
the same number of electrons.
A simplified version of the pseudoparticle methodology is well known already long time 
as slave-boson technique (see, e.g., Refs.~\cite{WingreenMeirPRB94,WingreenPRB96} - classics of quantum transport in junctions).
Recently, development of the dynamical mean field theory renewed interest in the
methodology.\cite{WernerRMP14}

In PP-NEGF central object of interest is single pseudoparticle Green function 
 \begin{equation}
 \label{defGPPNEGF}
 G_{S_1S_2}(\tau_1,\tau_2)=-i\left\langle T_c\,\hat p_{S_1}(\tau_1)\,\hat p_{S_2}^\dagger(\tau_2)\right\rangle
 \end{equation}
Contrary to the NEGF where intra-system interactions are treated by
the diagrammatic perturbation series, PP-NEGF is perturbative in system-baths couplings.
However, all the standard diagrammatic machinery (including methodology to build 
conserving approximations) of the NEGF is applicable also to pseudo-particle Green functions. 
The PP-NEGF is defined by solving the Dyson equation of the same structure as in Eq.(\ref{GDyson})
with the difference that self-energy (perturbatively) accounts for system-baths coupings. 
Green function (\ref{defGPPNEGF}) can be considered as a generalization of 
the reduced (system) density matrix $\sigma_{S_1S_2}(t)$. 
Indeed, while the latter provides information on  populations and coherences at a particular (local) time, 
PP-NEGF gives also temporal correlations; its lesser projection taken at equal times is 
the system density matrix:  $i\zeta_{S_1}G^{<}_{S_1S_2}(t,t)=\sigma_{S_1S_2}(t)$
(here $\zeta_S=+1$ ($-1$) for Bose (Fermi) state $\lvert S\rangle$).

The main technical difference between NEGF and PP-NEGF comes from the necessity 
to impose restriction (\ref{Q1}). This results in several unusual properties of pseudoparticle
Green functions. For example, NEGF fluctuation-dissipation relation, 
$G^{>}_{ij}(t_1,t_2)-G^{<}_{ij}(t_1,t_2)=G^{r}_{ij}(t_1,t_2)-G^{a}_{ij}(t_1,t_2)$,
becomes for the PP-NEGF in the $\hat Q=0$ subspace 
$G^{>}_{S_1S_2}(t_1,t_2)=G^{r}_{S_1S_2}(t_1,t_2)-G^{a}_{S_1S_2}(t_1,t_2)$ 
because $G^{<}_{S_1S_2}(t_1,t_2)$ does not have contributions in $\hat Q=0$;
Dyson equations for retarded ($\hat Q=0$ subspace) and lesser ($\hat Q=1$ subspace) 
projections are decoupled in the PP-NEGF;
to reflect physical reality any projection in the diagrammatic expansion
should contain only one lesser Green function (sum of charges $\hat Q$ from different contributions 
in any diagram should be $1$ - this is contribution from any lesser Green function, 
then the diagram as a whole belongs to physical subspace of 
the extended Hilbert space); etc. 
An interested reader is encouraged to consult Refs.~\cite{EcksteinWernerPRB10,OhAhnBubanjaPRB11,WhiteGalperinPCCP12,WernerRMP14}.
In particular, Ref.~\cite{EcksteinWernerPRB10} is a beautiful introduction to the methodology,
Ref.~\cite{WhiteGalperinPCCP12} contains explicit expresisions for self-energies due to 
coupling to fermionic (e.g., contacts) and bosonic (e.g., radiation field or thermal environment) baths.
  
To summarize, PP-NEGF has several important advantages:
1.~The method is conceptually simple; 
2.~Standard diagrammatic perturbation theory can be applied (in particular, this means that 
physical conservation laws are preserved within the methodology); 
3.~Already in its lowest (second) order in system-baths interactions, 
the non-crossing approximation (NCA), the pseudoparticle NEGF goes beyond standard
QME approaches by accounting for both non-Markovian effects and hybridization of molecular states; 
4.~The method is capable of treating transport in the language of many-body states of the isolated 
molecule, exactly accounting for all intra-molecular interactions. 
We stress that while (similar to the Liouville space formulation) PP-NEGF describes a system 
utilizing its many-body states (eigenstates of $\hat H_M$) and accounts (albeit perturbatively)
for hybridization between states of the system and baths,  
it avoids the two main problems of the Liouville space superoperator method
(as discussed at the end of Section~\ref{liouville}).  
As any approximate scheme the pseudoparticle NEGF has its own limitations,\cite{KotliarRMP06}
however those are important mostly at low temperatures (below Kondo temperature).
Also, lowest order of the method (NCA) was recently shown to be sensitive to details of 
accompanying approximations.\cite{CohenMillisReichmanPRB16} 
Finally, an important deficiency of the methodology is related to its inability to yield information on 
full counting statistics. Presumably, the problem comes from its formulation
within extended (unphysical) Hilbert space. We now turn to a methodology which allows to
overcome the difficulty. 

\subsubsection{Hubbard NEGF}\label{hubnegf}
Hubbard Green functions were originally introduced as a tool to develop perturbative
expansion about the atomic limit in interaction between atoms with the goal to describe 
electron correlations in  narrow energy bands.\cite{Hubbard1967} 
The methodology was further developed and applied to study magnetically ordered systems.\cite{IzyumovSkryabin_1988,IzyumovKatsnelsonSkryabin_1994,OvchinnikovValkov_2004}
These considerations were focused on equilibrium strongly correlated lattice models.  
Recently we utilized the approach to consider nonequilibrum atomic limit,
thus introducing diagrammatric technique for nonequilibrium Hubbard Green functions.\cite{ChenOchoaMGJCP17}

The Hubbard NEGF is capable of describing physics of open current carrying nanojunctions 
starting from atomic limit (system and baths are decoupled) with all intra-system interactions
taken into account exactly; system-baths couplings are used as small parameters in 
perturbative expansion. In this sense it is similar to PP-NEGF discussed above with
important difference that the methodology is formulated solely in physical Hilbert space.
Contrary to PP-NEGF, which studies temporal correlations between pairs of many-body states 
of the system, Eq.(\ref{defGPPNEGF}), Hubbard NEGF focuses on similar correlations 
between transitions from one many-body state to another. 
The latter are described by Hubbard (or projection) operators
\begin{equation}
\label{defX}
 \hat X_{S_1S_2} = \lvert S_1\rangle \langle S_2\rvert
\end{equation}
The correlation function (nonequilibrium Hubbard Green function) is defined on the contour as
\begin{equation}
\label{defGHubNEGF}
G_{(S_1S_2),(S_3S_4)}(\tau_1,\tau_2) = 
 -i\left\langle T_c\,\hat X_{S_1S_2}(\tau_1)\,\hat X_{S_3S_4}^\dagger(\tau_2)\right\rangle
\end{equation}
This definition is similar in spirit to the NEGF. Indeed, spectral decomposition of a quasiparticle 
annihilation operator, $\hat d_i=\sum_{S_1,S_2}\langle S_1\rvert\hat d_i\lvert S_2\rangle\,\hat X_{S_1S_2}$,
immediately shows connection between (\ref{defGNEGF}) and (\ref{defGHubNEGF}).
Indices $m$ in Eqs.~(\ref{VMP})-(\ref{IK}) are such transitions between many-body states,
$m=(S_1S_2)$:
Bose type transitions in Eqs.~(\ref{VMP}) and (\ref{Ip}) and Fermi - in Eqs.~(\ref{VMK}) and (\ref{IK}).

Cornerstone for both NEGF and PP-NEGF diagrammatic techniques is the Wick's theorem\cite{FetterWalecka_1971,Danielewicz1984}
which relies on (anti)commutation relations for creation and annihilation (Fermi) Bose operators:
$[\hat d_i;\hat d_j^\dagger]_\pm=\delta_{i,j}$ and 
$[\hat p_{S_1};\hat p_{S_2}^\dagger]_\pm=\delta_{S_1,S_2}$, respectively.
It is crucial that result of (anti)commutation is a number. This is not so for Hubbard
operators (\ref{defX}): $[\hat X_{S_1S_2};\hat X_{S_3S_4}^\dagger]_\pm = \delta_{S_2,S_4}\hat X_{S_1S_3}\pm\delta_{S_1,S_3}\hat X_{S_4S_2}$. 
Nevertheless, a variant of Wick's theorem for Hubbard operators was developed 
for equilibrium systems.\cite{IzyumovSkryabin_1988,IzyumovKatsnelsonSkryabin_1994,OvchinnikovValkov_2004}
The consideration was based on commutation properties of equilibrium density matrix with 
Hubbard operators. In junctions one has to deal with a mixture of quasiparticle excitations
in the baths and Hubbard operators describing eigenstates of the system Hamiltonian $\hat H_M$.
Moreover, the system is in a nonequilibrium state. 
To build nonequilibrium diagrammatic technique for Hubbard Green functions we
made two assumptions: 1.~originally (at $t_0\to -\infty$) system and baths were decoupled and 
the  system (molecule) was in thermal equilibrium and 2.~after coupling was established 
the system reached steady-state defined solely by bath induced boundary conditions (i.e. memory
of the initial state was lost). The latter is a usual assumption within the NEGF, and thus the former 
is unimportant for long time behavior of the system. Choice of thermal equilibrium
as the initial condition allows to employ the Wick's theorem of Refs.~\cite{IzyumovSkryabin_1988,IzyumovKatsnelsonSkryabin_1994,OvchinnikovValkov_2004}
for Hubbard operators, while the standard Wick's theorem\cite{FetterWalecka_1971,Danielewicz1984} 
is utilized to decouple quasiparticle correlation functions in perturbative expansion. 
The latter introduce boundary conditions imposed by baths on the system.
After expansion to desired order in the coupling is finished, the diagrams are dressed in complete 
analogy with the standard diagrammatic technique. This results in a modified version of 
the Dyson type equation
\begin{align}
\label{Dyson1}
& G_{m_1m_2}(\tau_1,\tau_2) = \sum_{m_3}\int_c d\tau_3\, g_{m_1m_3}(\tau_1,\tau_3)\, P_{m_3m_2}(\tau_3,\tau_2)
\\
\label{Dyson2}
& g_{m_1m_2}(\tau_1,\tau_2) = g^{0}_{m_1m_2}(\tau_1,\tau_2) 
\\ & \quad
+ \sum_{m_3,m_4}\int_c d\tau_3\int_c d\tau_4\, 
g^{0}_{m_1m_3}(\tau_1,\tau_3)\,\Sigma_{m_3m_4}(\tau_3,\tau_4)\,g_{m_4m_2}(\tau_4,\tau_2)
\nonumber
\end{align}
Here $m_i$ stands for transition between a pair of many-body states, 
$G_{m_1m_2}(\tau_1,\tau_2)$ is the Hubbard Green function (\ref{defGHubNEGF}),
$g_{m_1m_2}(\tau_1,\tau_2)$ is the locator, $g^{0}_{m_1m_2}(\tau_1,\tau_2)$ is the locator
in the absence of coupling to the baths, and $P_{m_1m_2}(\tau_1,\tau_2)$ is the strength operator.
Eqs.~(\ref{Dyson1})-(\ref{Dyson2}) are exact in the same sense as the usual Dyson equation.
For details of derivation and rules of the nonequilibrium diagrammatic technique interested
reader is encouraged to consult Ref.~\cite{ChenOchoaMGJCP17} and references therein. 

\begin{figure}[htbp]
\centering
  \includegraphics[width=\linewidth]{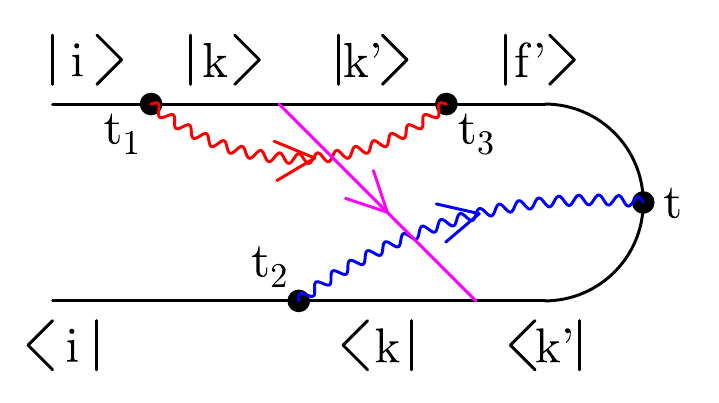}
  \caption{An example of optical scattering process in junction. $\lvert i\rangle$ and $\lvert k\rangle$
  are eigenstates of a neutral molecule, $\lvert k'\rangle$ and $\lvert f'\rangle$ are eigenstates of
  a cation. Wavy lines (red and blue) represent photon Green functions (\ref{defF}), 
  straight line (magenta) indicates electron self-energy due to coupling to contacts (\ref{defSK}). }
  \label{fig3}
\end{figure}

It is important to note that the nonequilibrium diagrammatic technique for Hubbard Green fucntions 
is a generalization of the Liouville superoperator formulation described in Section~\ref{liouville}.
On-the-contour diagrams are directly related to NEGF Feynman diagrams 
and account for both projections (as is the case in the double sided Feynman diagrams)  
and contractions between Hubbard operators of the system, quasiparticle operators representing electrons in contacts, and radiation field operators - 
respectively, Green functions (\ref{defGHubNEGF}), (\ref{defSK}), and (\ref{defF}) 
(compare Fig.~\ref{fig3} with Fig.~\ref{fig1}b).
Note that the Hubbard NEGF can also be considered as a Green function
generalization of the real-time perturbation theory developed for density matrices\cite{SchonPRB94,SchoellerSchonPRB96,Schoeller2000,LeijnseWegewijsPRB08}
and as an extension of auxiliary fields Hubbard Green function approach\cite{SandalovIJQC03,FranssonPRB05,SandalovJPCM06,SandalovPRB07,MGNitzanRatnerPRB08,Fransson_2010}
(as it yields an organized diagrammatic procedure to account for the system-baths couplings
and allows evaluation of multi-time correlation functions). 
Contrary to the PP-NEGF, the Hubbard NEGF is formulated solely in the physical Hilbert space.
As a result it can be utilized to study full counting statistics of transport 
(see preliminary data in Section~\ref{quantum} below).  
However, the approach (at current level of development) has an important formal limitation:
no clear way of constructing the Luttinger-Ward functional $\Phi$ has been proposed so far.
Thus, while model simulations which we performed in Ref.~\cite{ChenOchoaMGJCP17} 
show close correspondence with exact results, at the moment one cannot formally guarantee 
conserving character of the Hubbard NEGF diagrammatic expansions. 
A way to overcome the difficulty may be in constructing path integral formulation for the 
Hubbard NEGF utilizing generalized coherent states.\cite{Perelomov_1986}
This direction requires further research.
 
\section{Classical light}\label{classical}
Until recently most simulations of optical properties in junctions were performed utilizing
classical radiation fields. Roughly one can separate these studies into two groups: steady-state
and time-dependent considerations. In the latter group (and for harmonic driving) 
either transition to Floquet space,\cite{HoWangChuPRA86,GrossPRB08,RaiMGJPCC13,RabitzJCP14}
or transformation into the rotating frame of the field allow to formulate effective time-independent
problem.\cite{PeskinMGJCP12,GaoMGJCP16_1} 

\begin{figure*}[htbp]
\centering
  \includegraphics[width=\linewidth]{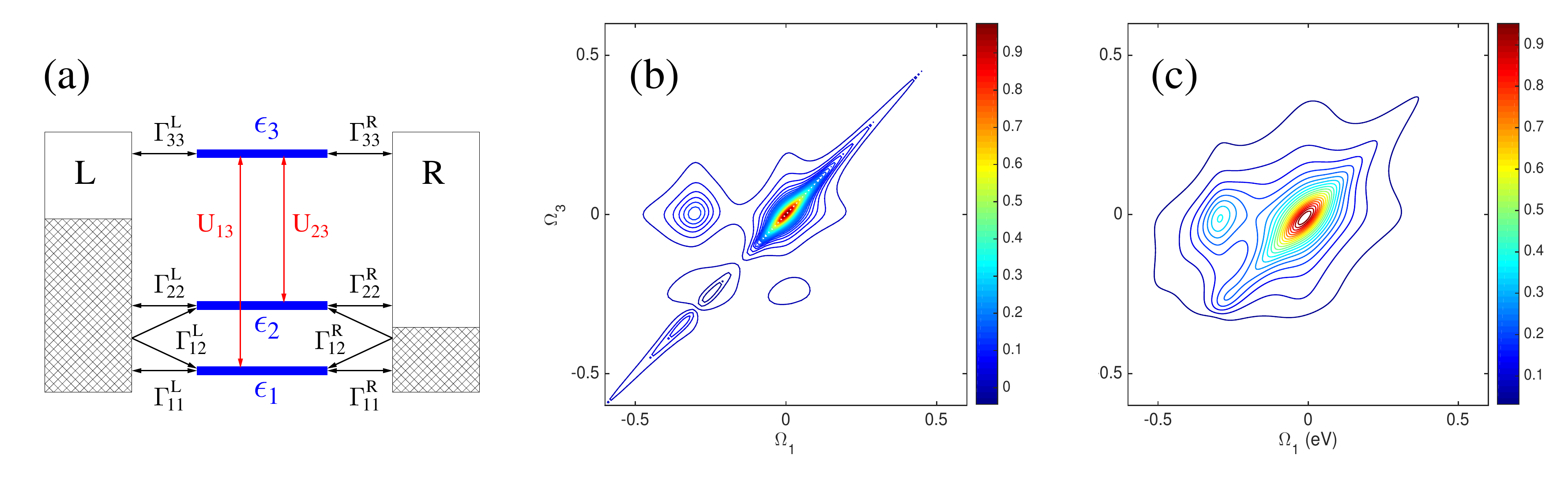}
  \caption{An example of 2D optical signal in a nanojunction. Shown are (a) junction model, 
  (b) NEGF (exact for the model) and (c) PP-NEGF results. 
  Lindblad/Redfied Liouville space formulation yields zero signal.
  Reprinted from [Y.~Gao and M.~Galperin, J. Chem. Phys., 2016, \textbf{144}, 244106],
  with the permission of AIP Publishing.
  }
  \label{fig4}
\end{figure*}

If light-matter coupling is relatively weak so that perturbative expansion in the interaction 
can be performed, bare PT expansion (as discussed in Section~\ref{liouville}) is justified, and 
standard tools of nonlinear optical spectroscopy can be used in studies of junctions.
These studies are often performed for steady-state regime (in the frequency domain). 
For example, Refs.~\cite{Harbola2006b,MukamelHarbolaJCTC15} utilized the Liouville space 
formulation to discuss current induced fluorescence in molecular junctions.
Ref.~\cite{HarbolaMukamelJCP14} studied stimulated and spontaneous light emission. 
Multidimensional optical spectroscopy in junctions was considered in Refs.~\cite{RahavMukamelJCP10,MukamelJCP15}.
Evaluation of the resulting multi-time electronic correlation functions, Eq.(\ref{rhoI}),
was performed either employing quasiparticle language and utilizing the standard \
Wick's theorem,\cite{Harbola2006b,RahavMukamelJCP10}
or relying on the quantum regression formula.\cite{HarbolaMukamelJCP14,MukamelHarbolaJCTC15,MukamelJCP15}
The former way is exact; the price to pay is necessity to work in the quasiparticle (orbital) basis
assuming noninteracting (quadratic) character of the molecular Hamiltonian. Such assumptions are
quite common in DFT based simulations, however one has to be cautious when taking
Kohn-Sham orbitals as proper representations for molecular orbitals. In particular, in junctions
the approach may lead to qualitative failures in predicting junction responses to external 
perturbations.\cite{BaratzMGBaerJPCC13} 
Possible pitfalls of the regression formula were discussed in Section~\ref{liouville}.
For example, in Ref.~\cite{GaoMGJCP16_2} we used a three level model to demonstrate 
that the regression formula (when quantum master equation utilizes second order
to account for system-baths couplings) fails to reproduce  coherent 2D optical response
of a junction. At the same time the PP-NEGF methodology (within the same, second order, 
level of treatment of system-baths couplings) yields qualitatively correct signal (see Fig.~\ref{fig4}).

\begin{figure}[htbp]
\centering
  \includegraphics[width=0.8\linewidth]{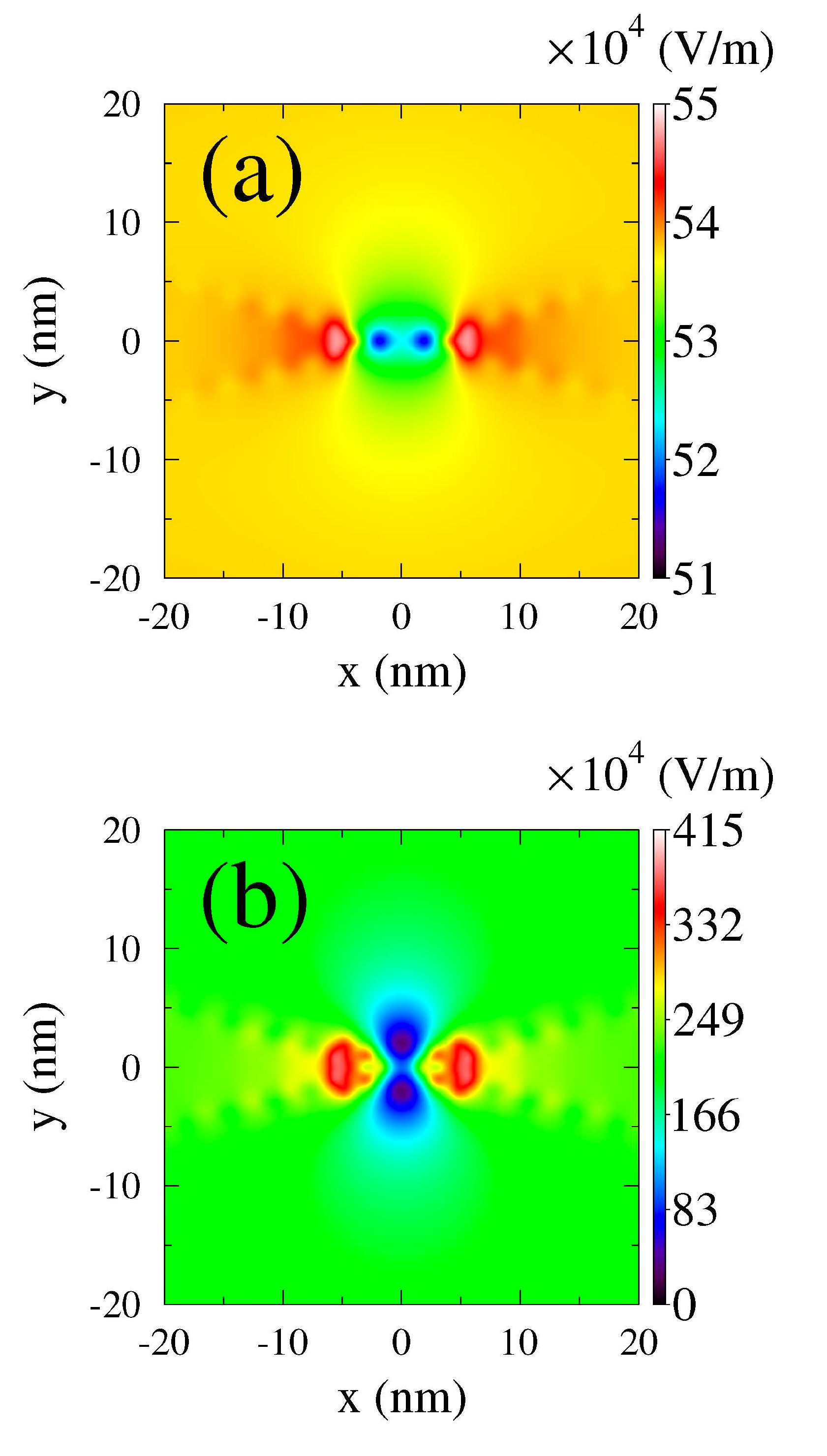}
  \caption{Instantaneous near field strength in a junction calculated 
  (a) without and (b) with the molecular response.
  Reprinted (figure) with permission from 
  [A.~.J.~White, M.~Sukharev, and M.~Galperin, Phys. Rev. B \textbf{86}, 205324, 2012.] 
  Copyright (2012) by the American Physical Society.
  http://dx.doi.org/10.1103/PhysRevB.86.205324
  }
  \label{fig5}
\end{figure}

\begin{figure}[t]
\centering
  \includegraphics[width=0.8\linewidth]{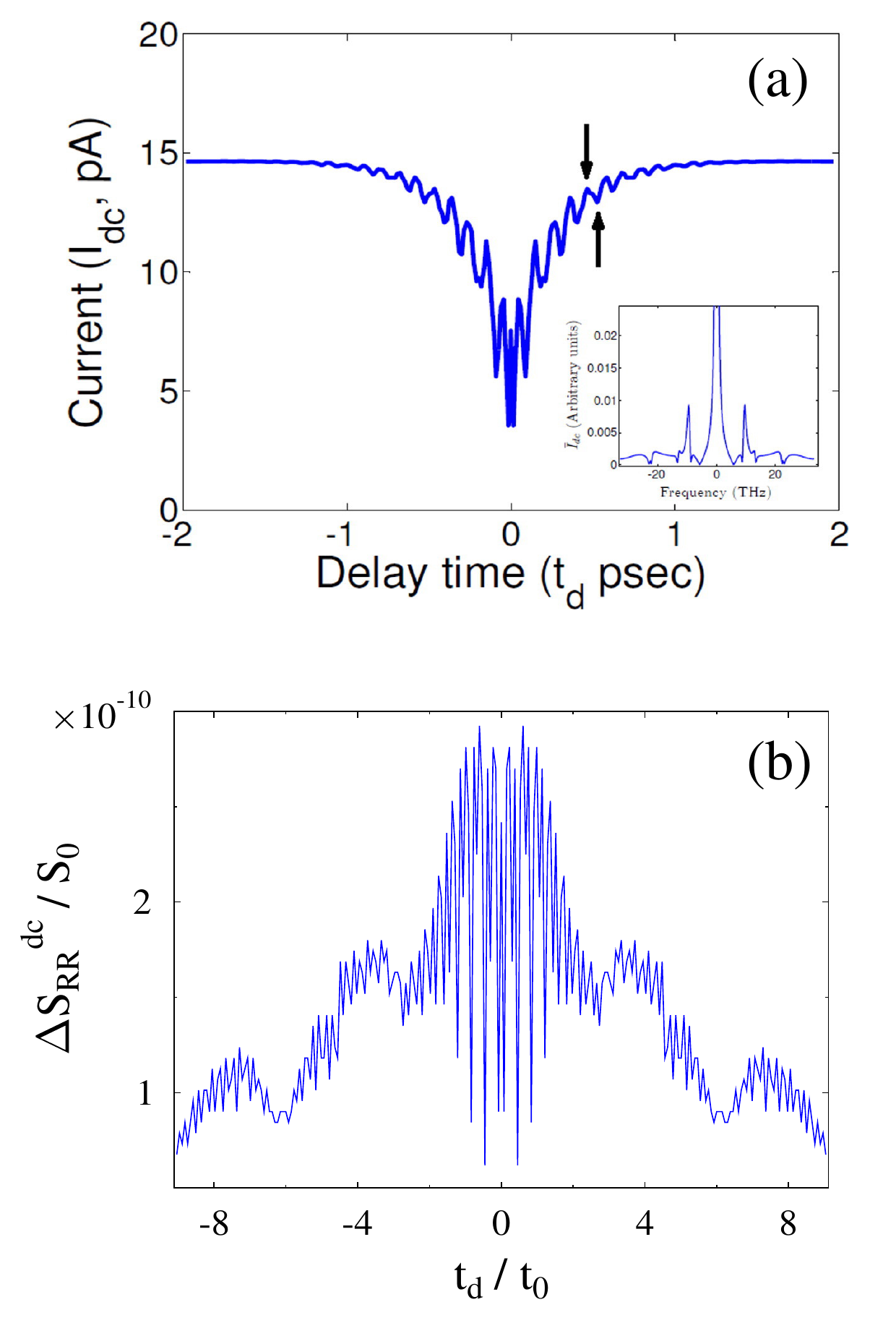}
  \caption{Pump-probe spectroscopy in nanojunctions. Laser pulse pair sequence
  induced (a) dc current and (b) dc noise plotted against delay time $t_d$
   reveal intra-molecular dynamics on the sub-pico-second time scale.
  Fig.~\ref{fig6}a reprinted with permission from 
  Y. Selzer and U. Peskin, J. Phys. Chem. C, 2013, \textbf{117}, 22369-22376.
  Copyright (2013) American Chemical Society.
  Fig.~\ref{fig6}b reprinted with permission from
  M.~A.~Ochoa, Y.~Selzer, U.~Peskin, and M.~Galperin, J. Phys. Chem. Lett., 2015, \textbf{6}, 470-476.
  Copyright (2013) American Chemical Society.
  }
  \label{fig6}
\end{figure}

Explicit time-dependent simulations with respect to spectroscopy in nanojunctions are often 
employed to simulate plasmon excitations induced in metallic contacts by external time-dependent  
radiation field. A numerical scheme propagates Maxwell equations
(e.g., the finite-difference time domain approach\cite{Gedney_2011} is a popular choice)
with the quantum system response entering the calculation via polarization current density.\cite{WhiteSukharevMGPRB12}
Junction dynamics is usually simulated within quantum master equation or Green function
approaches. A clear advantage of the former is time-locality of the density matrix; however,
low order treatments of system-contacts couplings may result in qualitative failures.\cite{EspMGPRB15,EspositoMGJPCC10}
An easy heuristic workaround is introduction of buffer zones,
which while being part of the dynamical calculation provide smooth connection between
nonequilibrium system and equilibrium baths.\cite{KronikHodJCTC14,HansenFrancoJPCC14}   
More rigorous (yet still not too heavy to remain practical) methodology is the hierarchical 
equation of motion approach;\cite{YanJCP08,CohenReichmanMillisPRB13,YanJCP14}
its main limitation is restriction to high temperatures.
Dynamical simulations employing Green functions are naturally more demanding,\cite{JauhoWingreenMeirPRB94,vanLeeuwenJCP09,LeeuwenJPCS10,EcksteinWernerPRB10}
and usually (when going beyond adiabatic regime) approximations are required to make 
the approach practical.\cite{LeeuwenStefanucciPRB14} 
An interesting development is a representation which maps
time-nonlocal interacting Dyson equation onto a noninteracting auxiliary Hamiltonian
with additional bath degrees of freedom; the latter problem can be efficiently solved.\cite{BalzerEcksteinPRB14} 
We employed NEGF within the wide-band approximation to simulate 
transport and optical response of molecular junctions driven by time-dependent plasmonic field.\cite{SukharevMGPRB10,FainbergSukharevParkMGPRB11,WhiteSukharevMGPRB12}
Among other things we demonstrated importance of molecular response in formation of the local field
(see Fig~\ref{fig5}) and hence also in junction responses .
The latter indicates that due to the crucial role of plasmonic enhancement in junction spectroscopies
one  has to be careful when applying bare PT treatments to study optics in nanojunctions
even with classical fields.  

\begin{figure}[t]
\centering
  \includegraphics[width=0.9\linewidth]{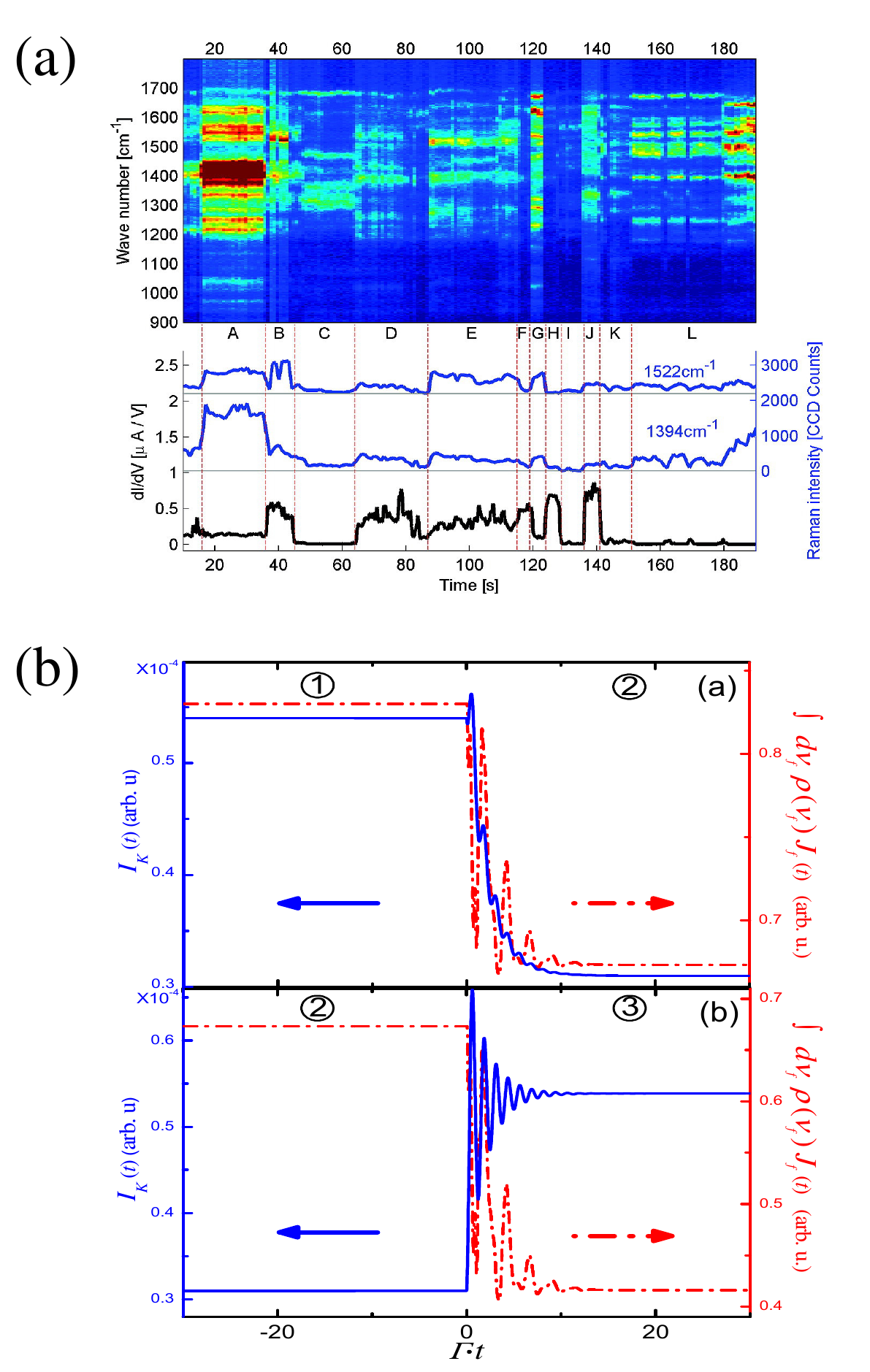}
  \caption{Time-dependent fluctuations in simultaneously measured electronic conductance and Raman response in molecular junctions. Shown are (a) experimental data and (b) a theoretical analysis.  
Fig.\ref{fig7}a reprinted with permission from D.~R.~Ward, N.~J.~Halas, J.~W.~Ciszek, 
J.~M.~Tour, Y.~Wu, P.~Nordlander, D.~Natelson, Nano Lett. 2008, \textbf{8}, 919-924.
Copyright (2008) American Chemical Society.
Fig.~\ref{fig7}b reprinted with permission from [T.-H.~Park and M.~Galperin, Phys. Rev. B, \textbf{84}, 075447 (2011).] Copyright (2011) by the American Physical Society.
http://dx.doi.org/10.1103/PhysRevB.84.075447
  }
  \label{fig7}
\end{figure}

Another case where working in time-domain may be preferable is pump-probe type spectroscopy. 
Quantitative mapping of fast voltage pulse by plasmonic luminiscence (probe) 
was demonstrated in STM junction measurements.\cite{LothAPL13}
An opposite proposal of pumping by light (laser pulse pairs sequences) 
and probing dc current\cite{SelzerPeskinJPCC13} and noise\cite{OchoaSelzerPeskinMGJPCL15} 
was put forward as a way to access intra-molecular dynamics on the sub-picosecond time scale.\footnote{DC transport measurements are an important part of the suggestion,
because electronic components are too slow to directly measure dynamics with picosecond resolution.} 
Theoretical simulations were performed utilizing
quantum master equation\cite{SelzerPeskinJPCC13} and NEGF\cite{OchoaSelzerPeskinMGJPCL15}
(see Fig.~\ref{fig6});
experimental verification of the proposed approach is an ongoing research
in the group of Prof. Yoram Selzer.

Finally, a distinct feature of spectroscopy in open systems (nanojunctions)
is the fact that photons and electrons participate in the same scattering process
but contribute to different separately measured signals (e.g., photon flux and electron current).
This simple idea is behind all the suggestions of measuring transport characteristics of one agent 
to describe properties of the other. 
In terms of theoretical treatment this is an indication of necessity to treat both
photon and electron fluxes, Eqs.~(\ref{Ip}) and (\ref{IK}), at the same level of theory.
Technically this is seen from the fact that both fluxes can be expressed 
in terms of the same correlation (Green) functions, which has to be evaluated at the same level of 
theory in both cases. For example, Refs.~\cite{ParkMGEPL11,ParkMGPRB11} consider 
temporal correlations between Raman signal and conductance for a model of junction
driven by time-dependent classical field (see Fig.~\ref{fig7}). 
Electron Green function, which enters expressions for both fluxes, was evaluated exactly 
with respect to light-matter coupling.


\begin{figure*}[htbp]
\centering
  \includegraphics[width=\linewidth]{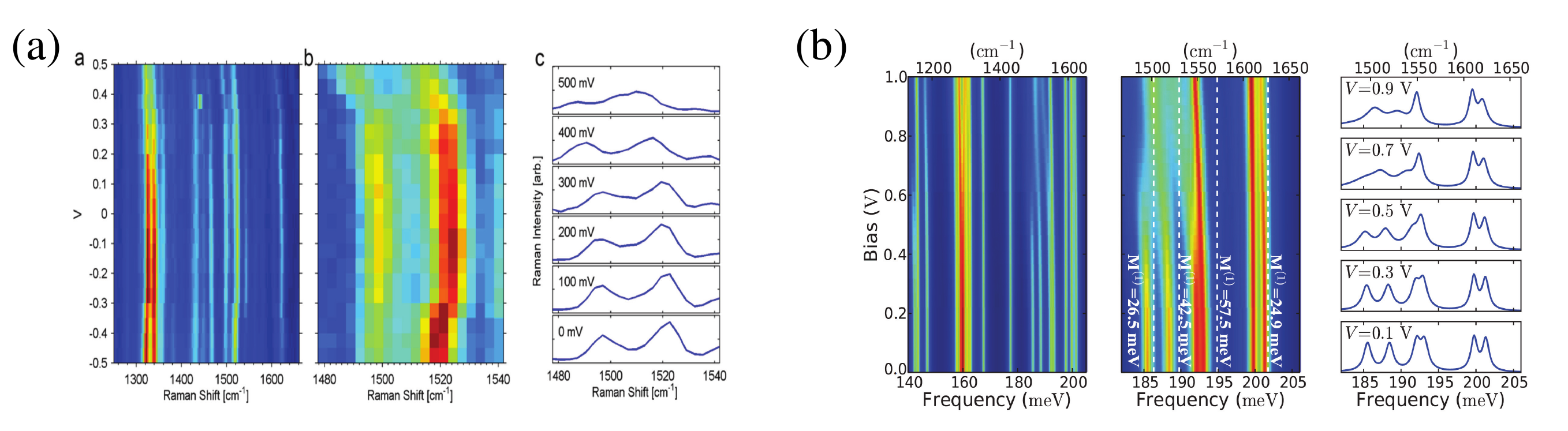}
  \caption{Bias dependence of Raman shift in OPV3 junction. 
  Shown are (a) experimental data and (b) a theoretical analysis.  
  Fig.~\ref{fig8}a reprinted by permission from Macmillan Publishers Ltd: [Nature Nanotechnology]
  (Ref.~\cite{NatelsonNatNano11}), copyright (2011).
  Fig.~\ref{fig8}b reprinted with permission from [K.~Kaasbjerg, T.~Novotn{\' y}, and A.~Nitzan,
  Phys. Rev. B \textbf{88}, 201405 (2013).] Copyright (2013) by the American Physical Society.
  http://dx.doi.org/10.1103/PhysRevB.88.201405
  }
  \label{fig8}
\end{figure*}

\section{Quantum light}\label{quantum}
Quantum treatment of radiation field is required when purely quantum effects are pronounced in 
the light-matter interaction. These effects include photons in entangled and squeezed states,
measurements of counting statistics of photon flux, interactions
induced by quantum light in electron subsystem, etc. Note that the latter are present even in 
the absence of an external field.\cite{BaratzWhiteMGBaerJPCL14}

Turning to theoretical treatment of Raman spectroscopy in nanojunctions,
it is worth mentioning that any spontaneous light emission (SLE) has to be considered
quantum mechanically.\cite{Mukamel_1995} 
Thus, corresponding theoretical derivations always start from 
quantum treatment of the field, Eq.(\ref{Ip}). After the derivation is completed,
one can switch to classical description. Then bare PT and corresponding multi-time correlation
functions, Eq.(\ref{rhoI}), become a safe way for description of optical response in junctions.
Alternatively one may decide to stay with quantum treatment of the field.
This is what was done in our NEGF and PP-NEGF theoretical studies of Raman spectroscopy 
in junctions.\cite{GalperinRatnerNitzanNL09,GalperinRatnerNitzanJCP09,MGANJPCL11,MGANPRB11,OrenMGANPRB12,WhiteTretiakNL14,GaoMGANJCP16,ApkarianMGANPRB16}
Following the standard nonlinear optical spectroscopy formulation, they rely on bare PT 
expansion in the light-matter interaction.
Resulting treatment is to some extent similar to that of 
Refs.~\cite{Harbola2006b,RahavMukamelJCP10} for NEGF treatments and
is better than state-based formulations of Refs.~\cite{HarbolaMukamelJCP14,MukamelHarbolaJCTC15,MukamelJCP15} 
in treating molecule-contacts coupling for PP-NEGF treatments;
the only difference is that Green function approaches make it easier to separate specifically
Raman diagrams from other SLE contributions.\cite{MGRatnerNitzanJCP15}
The central point is that these are bare PT (in light-matter coupling) considerations,
and as such they describe light scattering from broadened (due to molecule-contacts hybridization)
nonequilibrium current-carrying states (or levels) of the molecule. This is not a comprehensive
treatment of responses of an optoelectronic device. 
Nevertheless, such approaches are still useful for qualitative understanding of Raman scattering in nanojunctions. 

\begin{figure}[htbp]
\centering
  \includegraphics[width=0.8\linewidth]{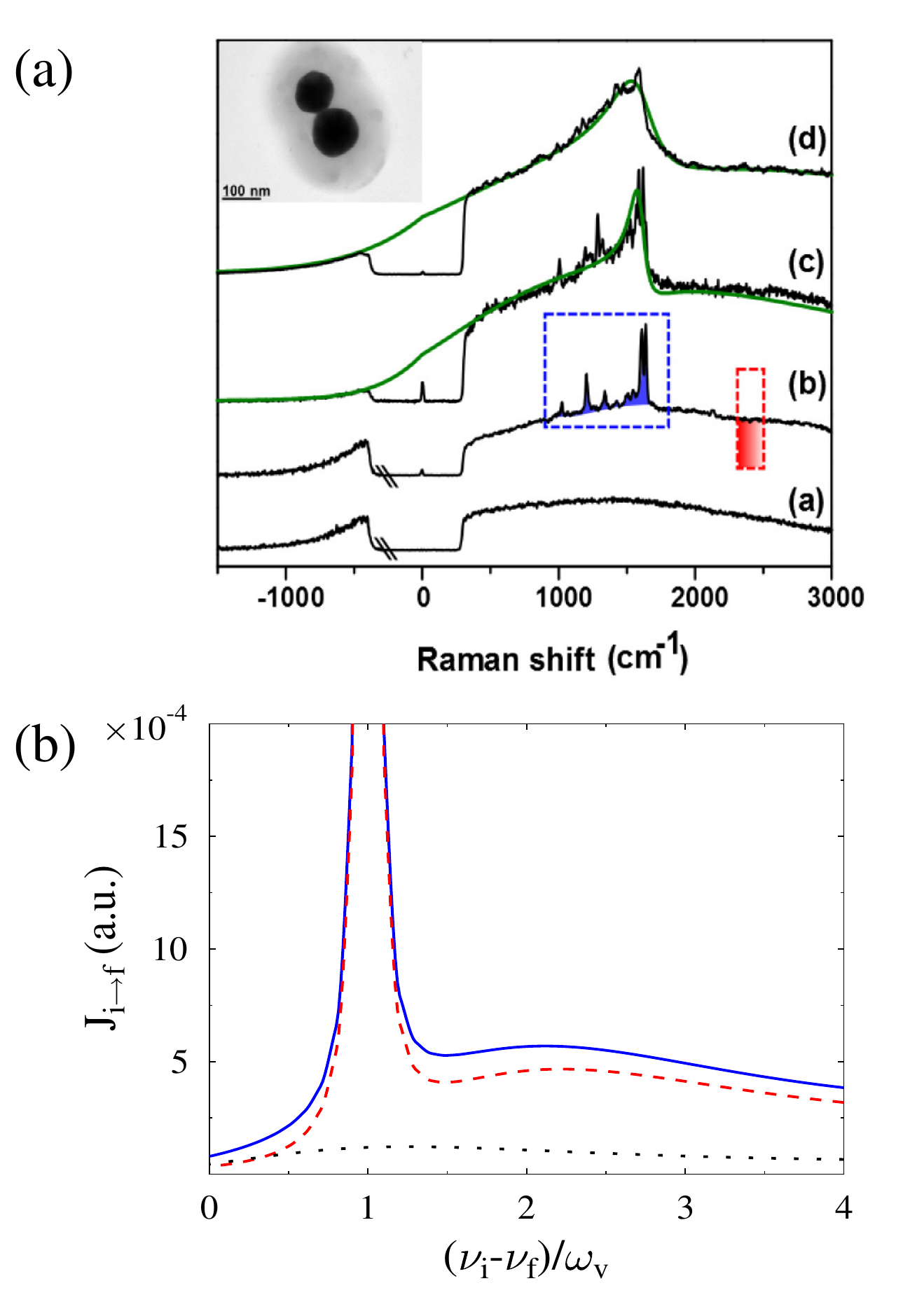}
  \caption{Fano-like lineshapes in the Raman spectra of molecules adsorbed at metal interfaces:
  (a) experimental data and (b) a theoretical analysis. Plotted in (b) are pure electronic (black dotted line),
  electronic-dressed vibrational Raman (red dashed line), and their sum (full blue line).
  Figure reprinted with permission from [S.~Dey, M.~Banik, E.~Hulkko, K.~Rodriguez, 
  V.~A.~Apkarian, M.~Galperin, and A.~Nitzan, Phys. Rev. B \textbf{93}, 035411 (2016).]
  Copyright (2016) by the American Physical Society.
  http://dx.doi.org/10.1103/PhysRevB.93.035411
  }
  \label{fig9}
\end{figure}

We now give a short overview of recent theoretical studies of Raman scattering 
in current carrying junctions. 
After first simultaneous measurements of Raman and conductance in
molecular junctions were reported,\cite{CheshnovskySelzerNatNano08,NatelsonNL08} 
a theory of Raman scattering from current carrying molecular states was developed  
(along the lines discussed above) in
Refs.~\cite{GalperinRatnerNitzanNL09,GalperinRatnerNitzanJCP09} and utilized
to discuss estimation of a `nonequilibrium temperature' of molecular vibrations 
(extent of heating of the vibrations by electron flux) from ratio of Stokes and anti-Stokes
peak intensities.
Refs.~\cite{MGANJPCL11,MGANPRB11} extended the latter analysis to electronic
heating in molecular junctions in an attempt to interpret measurements presented 
in Ref.~\cite{NatelsonNatNano11}.
It was found that contrary to vibrational heating in junctions, data on
electronic heating is much less reliable except at very low biases.
In particular, modeling showed that the main contribution observed in the
experiments as electronic heating may result from non-equilibrium electronic
distribution in the molecule, while contribution from actual electronic heating in the
contacts is negligible.

Charge transfer (chemical) contribution to surface enhanced Raman was discussed 
in Ref.~\cite{OrenMGANPRB12}. Here quantum bare PT treatment of light-matter interaction 
was compared with quasi-classical approach. The latter was shown to be inadequate 
at biases beyond threshold defined by characteristic frequencies of molecular
vibrations (i.e. when inelastic effects become pronounced).

Experimentally observed bending of Stokes lines under bias in an OPV3 junction\cite{NatelsonPCCP13}
was explained by dependence of molecular vibrational normal modes
on charging state of the molecule in Ref.~\cite{KaasbjergNitzanPRB13}.
Charging induced frequency renormalization was studied in Ref.~\cite{KaasbjergNitzanPRB13} 
using a model where molecule-phonon coupling was taken into account up 
to quadratic term in shift from equilibrium (this quadratic term yields anharmonic effects in the model). 
Estimating model parameters from first principle simulations resulted in shift of vibrational 
frequency similar to experimental data on Stokes line bending (see Fig.~\ref{fig8}).
Utilization of state-based approach (the PP-NEGF) in first principles simulation\cite{WhiteTretiakNL14}
provided an easy way to simulate the Raman spectrum.
Studies of charge-induced renormalization of vibrational frequencies
were later reported also in Refs.~\cite{KronikNeatonNatelsonPNAS14,KronikNeatonNatelsonNL16}.

\begin{figure}[t]
\centering
  \includegraphics[width=\linewidth]{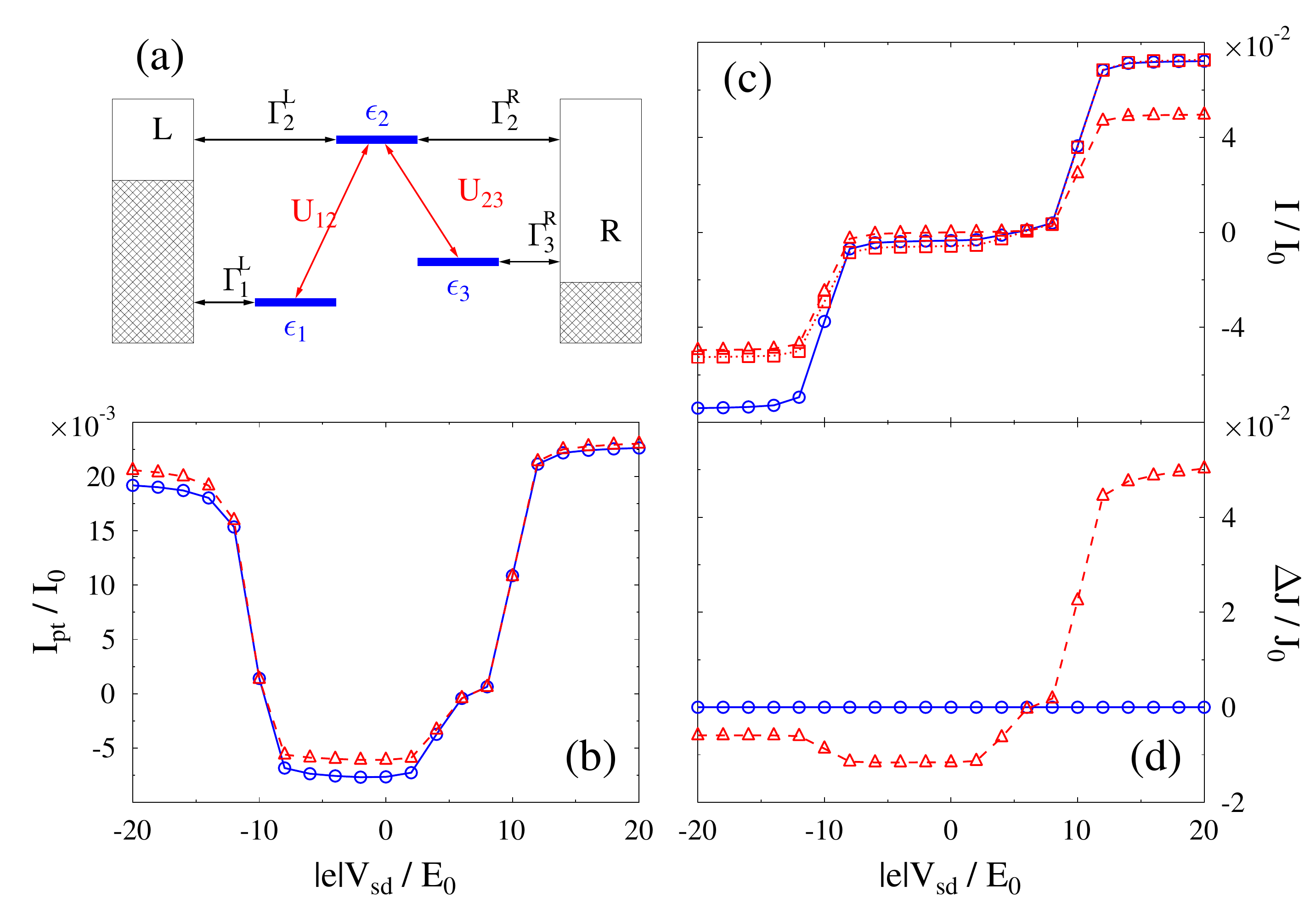}
  \caption{Physical conservation las in junction spectroscopy.
  For a three-level model (a) calculations of photon (b) electron (c) and energy (d) fluxes within
  diagrammatic (solid blue line) and bare (dashed and dotted red lines) perturbation theories
  show violation of conservation laws by the latter.
  Reprinted from [Y.~Gao and M.~Galperin, J. Chem. Phys. 2016, \textbf{144}, 174113],
  with the permission of AIP Publishing.
  }
  \label{fig10}
\end{figure}

Besides Raman shift and intensity of Stokes and anti-Stokes lines, widths and shapes of Raman peaks 
may also be a source of information on junction structure. Width of Stokes line and its dependence on
junction characteristics  (relaxation rates, proximity of electronic level to Fermi energy, and bias)
were discussed within a generic HOMO-LUMO model in Ref.~\cite{GaoMGANJCP16}.
Experimental observation and theoretical analysis of Fano-like lineshapes in the Raman
spectra was presented in Ref.~\cite{ApkarianMGANPRB16} (see Fig~\ref{fig9}).
The study found that observed Fano-like features in principle can be given by
interference between vibrational and electronic Raman scattering amplitudes (the Fano resonance);
however model calculations suggested that the observed lineshape asymmetry was dominated 
by purely electronic scattering sidebands that dress vibrational Stokes peaks.

\begin{figure}[t]
\centering
  \includegraphics[width=0.8\linewidth]{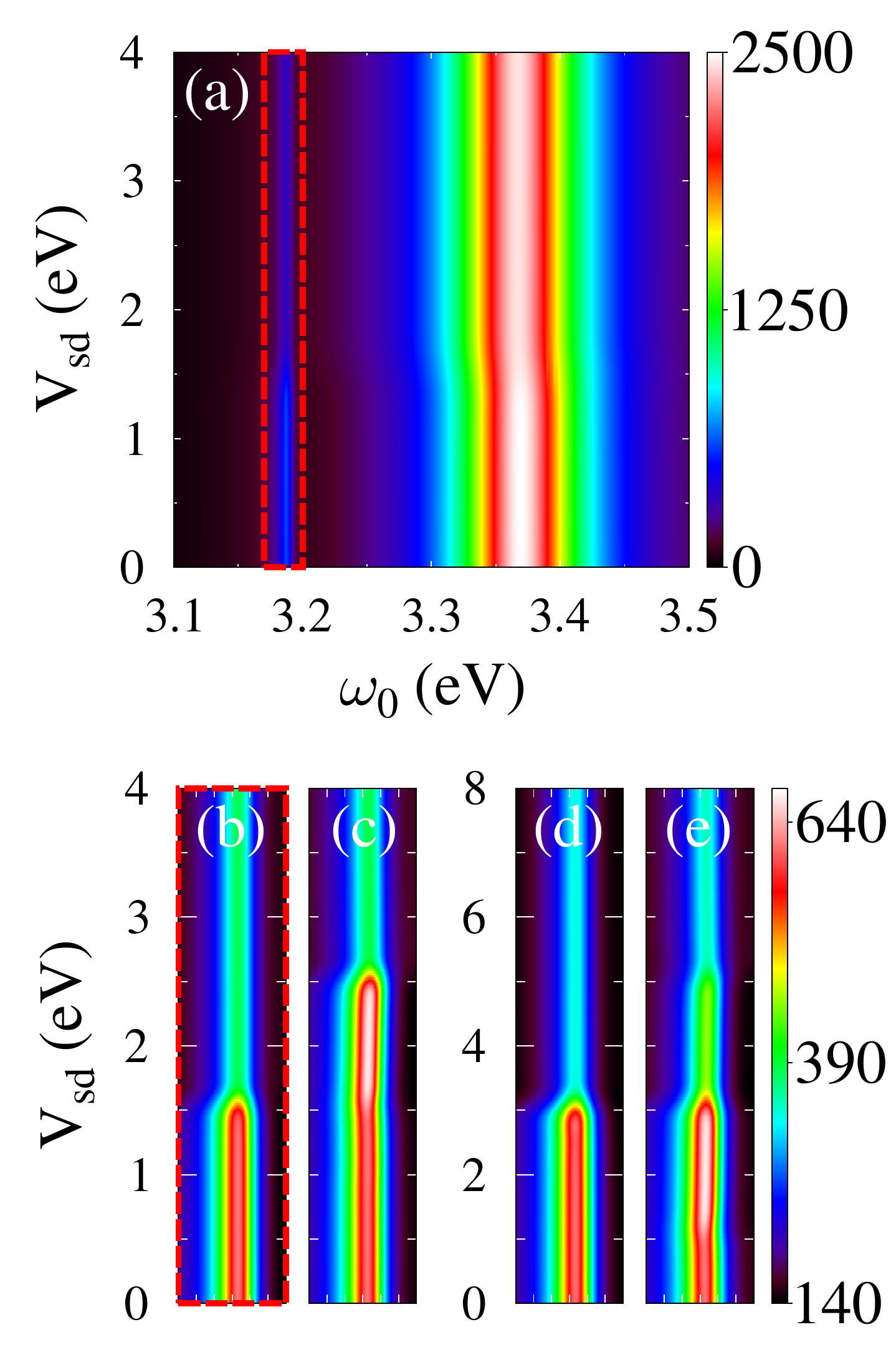}
  \caption{Strong exciton-plasmon coupling in junctions.
  PP-NEGF study of plasmon absorption spectrum as a function of bias (a)
  and close-up of Fano resonance (b-e) for different intra-system interactions
  and bias profiles.
  Reprinted with permission from (A.~J.~White, B.~D.~Fainberg, and M.~Galperin,
  J. Phys. Chem. Lett. 2012,  \textbf{3}, 2738-2743).
  Copyright (2012) American Chemical Society.
  }
  \label{fig11}
\end{figure}

\begin{figure*}[t]
\centering
  \includegraphics[width=\linewidth]{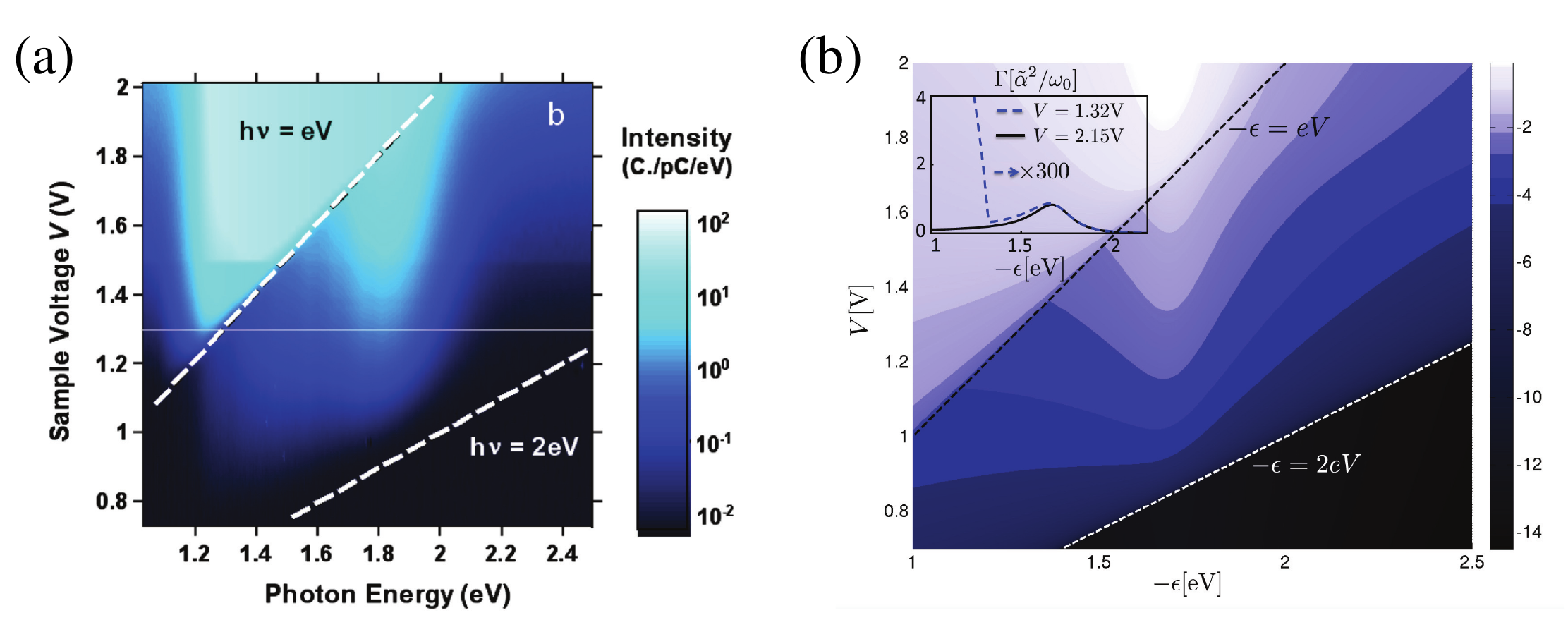}
  \caption{Bias induced light emission as a source of information on 
  electron-electron and electron-plasmon interactions in junctions.
  Emission spectrum vs. applied bias: (a) experimental data and (b) a theoretical
  analysis.
  Fig.~\ref{fig12}a reprinted with permission from [G.~Schull, N.~Neel, P.~Johansson, and R.~Berndt,
  Phys.~Rev.~Lett. \textbf{102}, 057401 (2009).] Copyright (2014) by the American Physical Society.
  http://dx.doi.org/10.1103/PhysRevLett.102.057401
  Fig.~\ref{fig12}b reprinted with permission from [F.~Xu, C.~Holmqvist, and W.~Belzig,
  Phys. Rev. Lett. \textbf{113}, 066801 (2014).] Copyright (2014) by the American Physical Society.
  http://dx.doi.org/10.1103/PhysRevLett.113.066801
  }
  \label{fig12}
\end{figure*}

\begin{figure}[htbp]
\centering
  \includegraphics[width=\linewidth]{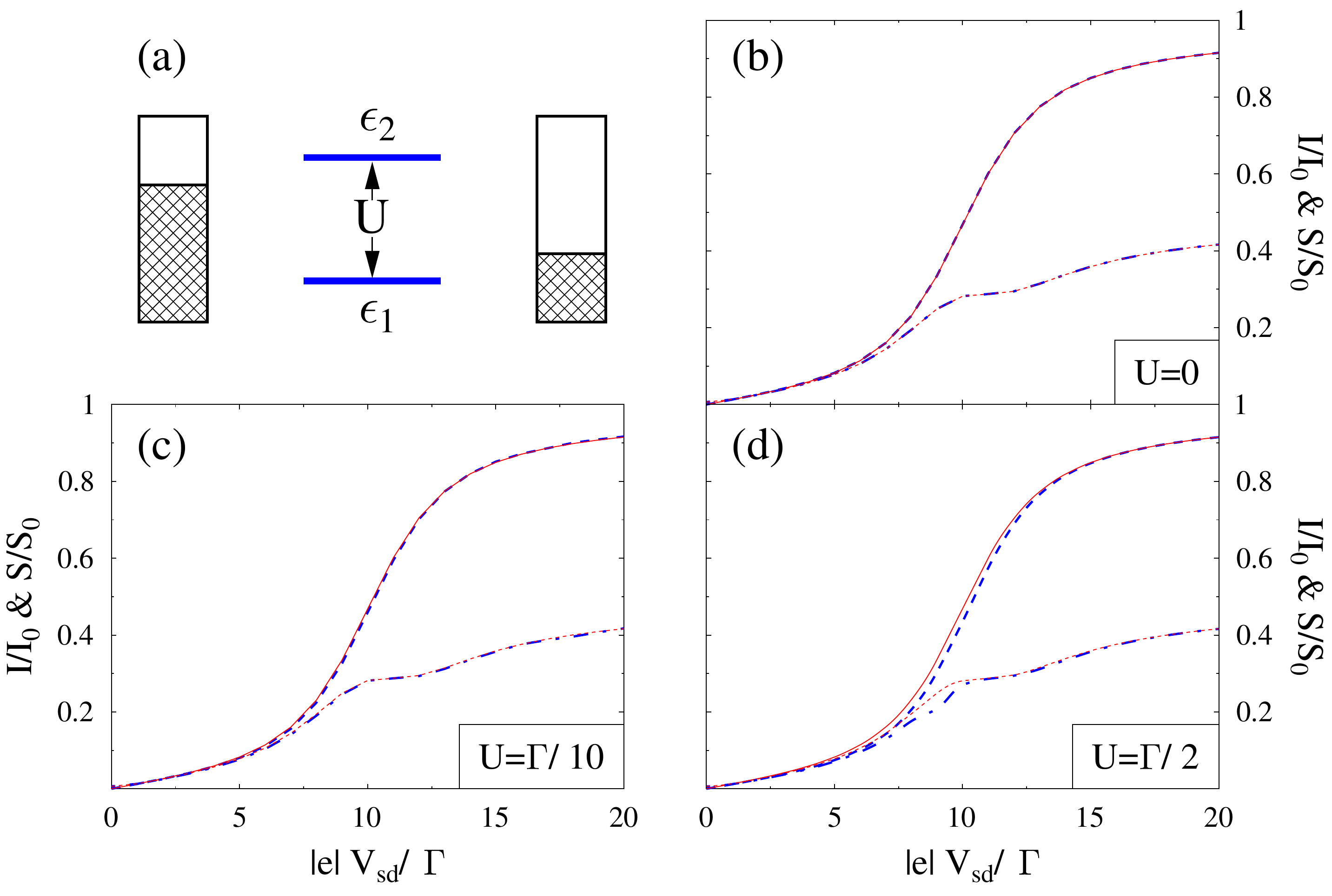}
  \caption{Full counting statistics of electron transport in junctions.
   Simulation is performed for Hubbard model (a),
   $\hat H_M=\sum_{i=1,2}\epsilon_i\hat n_i+U\hat n_1\hat n_2$,
   with parameters $\epsilon_2=-\epsilon_1=5\Gamma$.  
   Shown are current ($I_0=e\Gamma/\hbar$, solid red and dashed blue lines)
   and zero frequency noise ($S_0=e^2\Gamma/\hbar$, dotted red and dash-dotted blue lines)
   calculated, respectively,  within NEGF (red) and Hubbard NEGF (blue) methodologies
   for (b) $U=0$ (NEGF is exact here), (c) $U=\Gamma/10$, and (d) $U=\Gamma/2$. 
   In (c) and (d) NEGF utilizes second order diagrammatic perturbation theory in $U$,
   Hubbard NEGF is second order in system-baths coupling (i.e. first order in $\Gamma$).\cite{tbp}}
  \label{fig13}
\end{figure}

As stated above a comprehensive treatment of an optoelectronic device
should account for all the fluxes (photon and electron) at the same level of theory.
When considering quantum field such comprehensive consideration requires especial care.
In particular (as was discussed in Section~\ref{negf}), for the consideration to satisfy 
physical conservation laws one is forced to abandon the bare PT approach. 
The latter is well known to be a non-conserving approximation which may fail qualitatively 
when vertical flow (redistribution of electronic population in energy) is present.\cite{BaymKadanoffPR61,BaymPR62,KadanoffBaym_1962,Haussmann_1999,StefanucciVanLeeuwen_2013}
For example, in the theory of inelastic transport, which is technically equivalent
to electron-photon interaction, second order bare PT (Born approximation)
is non-conserving, and one has to employ the self-consistent Born approximation to get
meaningful results.\cite{ParkMG_FCS_PRB11} 
The same situation holds for any other interaction which causes vertical flow in the system
(see, e.g., Ref.~\cite{vanLeeuwenPRB09}).
Thus, e.g., it is not surprising that any consideration of current-induced fluorescence with
radiation field being treated quantum mechanically leads to a self-consistent treatment.\cite{GalperinNitzanPRL05,GalperinNitzanJCP06}

The main difference between classical and quantum fields 
(with respect to conserving character of approximation) is ability of the latter to mediate photon 
supported effective electron-electron interaction. Technically this interaction comes in the form 
of electronic self-energy due to coupling to radiation field, which being approximated in 
an inappropriate way (e.g., within bare PT) leads to violation of charge and energy conservation laws.
A proper way to derive conserving approximations was discussed in Section~\ref{negf}
(see Eqs.~(\ref{SpPhi}) and (\ref{PiPhi} there).
We note that the restrictions on application of standard tools of nonlinear optical spectroscopy 
to nanojunctions are relevant only for radiation fields treated quantum mechanically,
because classical fields do not induce time-nonlocal interactions in electronic subsystem
(i.e. technically they do not produce self-energies).
To illustrate the point in Ref.~\cite{GaoMGJCP16_1} we utilized NEGF to consider 
optical and electronic responses (fluxes) of a nanojunction within the bare PT 
(as accepted in nonlinear spectroscopy community) and diagrammatic PT 
(as is usual in quantum transport considerations) approaches (see Fig.~\ref{fig10}). 

Quantum treatment of radiation field is also required for strong light-matter interaction.
Here light and matter degrees of freedom cannot be separately distinguished,
and thus theoretical treatment should be performed in the basis of eigenstates
of the Hamiltonian $\hat H_M+\hat H_p+\hat V_{MP}$, Eq.(\ref{H}),
accounting for the matter, the light, and interaction between them.
We note that Green function methods presented in Sections~\ref{ppnegf}
and \ref{hubnegf} are ideally suited for the task.
Note that also here conserving character of resulting approximation should be satisfied;
however this time it is self-energy due to coupling to other baths
(e.g., contacts or thermal environment, etc.) rather than self-energies coming from accounting
for the light-matter interaction, are to be build properly.
In Ref.~\cite{WhiteFainbergMGJPCL12} we utilized the PP-NEGF to study strong 
molecule-plasmon interaction in  nanojunctions. Fig.~\ref{fig11} shows sensitivity of 
a molecule-plasmon Fano resonance to junction bias and intra-molecular interactions.

Finally, theoretical treatments related to statistics of photon flux also require 
quantum description of the radiation field. 
Note that an accurate (conserving, as discussed above) treatment in this case
is even more important: while in Ref.~\cite{GaoMGJCP16_1} we demonstrated sensitivity of 
flux (first cumulant of the full counting statistics, FCS) higher order cumulants 
of the FCS are much more sensitive to details of theoretical modeling.\cite{EspMGPRB15,NitzanPRB16} 
Participation of photons and electrons in the same process 
reveals itself in inter-dependence of optical and transport characteristics of an optoelectronic device. 
For example, ability of the plasmon emission spectrum 
to characterize finite frequency quantum noise of electron transport was demonstrated 
experimentally\cite{BerndtPRL12} and discussed theoretically\cite{KaasbjergNitzanPRL15}
(see Eqs.~(10) and (12) in Ref.~\cite{KaasbjergNitzanPRL15} for formal connection between 
plasmonic light emission and the quantum noise).
Similarly, optical spectra as a source of information on multielectron processes in junctions
was measured\cite{BerndtPRL09} and studied theoretically\cite{BelzigPRL14} (see Fig.~\ref{fig12}).
Refs.~\cite{BrandesPRL07,BrandesPRB08} considered noise of 
photon
and electron fluxes as well as cross-correlation counting statistics. 
The consideration utilized quantum master equation within the Born-Markov approximation,
which is known to be problematic for description of relatively strong (compared to $k_BT$)
system-bath couplings $\Gamma$.\cite{EspGalpPRB09,EspositoMGJPCC10,EspMGPRB15}
For a molecule chemisorbed on metallic surface $\Gamma\sim 0.01-0.5$~eV\cite{KinoshitaJCP95,DattaKubiakJCP98}
while room temperature is $k_BT\sim 0.03$~eV. Thus Green function approaches, which are not 
limited by high temperature restriction, are preferable for treatment of counting statistics in
molecular junctions. From the two state-based Green function methodologies 
introduced above (Setcions~\ref{ppnegf} and \ref{hubnegf}) only the Hubbard NEGF provides 
possibility to simulate FCS (see Fig.~\ref{fig13}). Application of the Hubbard NEGF to
description of optoelectronic devices is a direction for future research. 

\section{Conclusions}\label{conclude}
In recent years optical experiments in current-carrying nanojunctions became a reality
indicating emergence of a new field of research coined optoelectronics. 
Experimental advances challenged theory to develop adequate approaches
to characterize responses of open nonequilibrium systems to external drivings.
The field of optoelectronics is a natural meeting point of (at least) two research communities:
nonlinear optical spectroscopy and quantum transport. Each of the communities has its own 
theoretical toolbox. We reviewed recent progress in the field
comparing theoretical treatments of optical spectroscopy in nanojunctions.
In particular, bare pertubation theory usually performed in the Liouville space
and formulated in the language of superoperators (a standard theoretical tool
in spectroscopic studies of isolated systems) was compared with  theoretical approaches 
accepted in quantum transport community. With respect to the latter we focus on 
the Hilbert space Green function based considerations. 
Standard nonequilibrium Green function (NEGF) was discussed together with 
its state-based flavors: pseudoparticle and Hubbard NEGF.
We argued that the Green function considerations yield a convenient tool
for optoelectronics when radiation field is treated either classically or quantum mechanically. 
We showed that bare perturbation theory becomes inapplicable, when
a comprehensive treatment of nanojunction responses to bias and quantized radiation field
is the goal of the study.
We conclude that the Hubbard NEGF is a promising methodology which
generalizes the standard tools of nonlinear optical spectroscopy,
which is capable of comprehensive studies of optoelectronic devices.
Further development of the methodology and its applications to nanojunctions spectroscopy
are directions for future research.

\section*{Acknowledgements}
This review is based upon work supported by the National Science Foundation under 
grant No. CHE-1565939 and the U.S. Department of Energy under grant No. DE-SC0006422.





\providecommand*{\mcitethebibliography}{\thebibliography}
\csname @ifundefined\endcsname{endmcitethebibliography}
{\let\endmcitethebibliography\endthebibliography}{}

\end{document}